\documentclass[12pt,preprint]{aastex}
\begin{document}

\received{}
\accepted{}

\slugcomment{ Printed \today}

\title{
Proper Motions of the Jets in the Region of HH~30 and HL/XZ Tau. 
Evidence for a Binary Exciting Source of the HH 30 Jet
}

\shorttitle{Proper Motions in HH~30 and HL/XZ Tau}

\author{
Guillem Anglada\altaffilmark{1},
Rosario L\'opez\altaffilmark{2},
Robert Estalella\altaffilmark{2},
Josefa Masegosa\altaffilmark{1},
Angels Riera\altaffilmark{2,3},
and
Alejandro C. Raga\altaffilmark{4} 
}

\altaffiltext{1}{
Instituto de Astrof\'{\i}sica de Andaluc\'{\i}a, CSIC,
Camino Bajo de Hu\'etor 50, E-18008 Granada, Spain;
guillem@iaa.es, pepa@iaa.es
}
\altaffiltext{2}{
Departament d'Astronomia i Meteorologia, Universitat de Barcelona,
Av.\ Diagonal 647, E-08028 Barcelona, Spain;
rosario.lopez@am.ub.es, robert.estalella@am.ub.es
}
\altaffiltext{3}{
Departament de F\'{\i}sica i Enginyeria Nuclear,
Universitat Polit\`ecnica de Catalunya, Vilanova i la Geltr\'u, Spain;
angels.riera@upc.es
}
\altaffiltext{4}{
Instituto de Ciencias Nucleares, UNAM, Apdo.\ Postal 70-543, 
04510 M\'exico D.F., Mexico;
raga@nucleares.unam.mx
}


\begin{abstract}
We present [SII] images of the HH~30 and HL/XZ~Tau region obtained at two
epochs, as well as long-slit optical spectroscopy of the HH~30 jet.  We
measured proper motions of $\sim$100--300~km~s$^{-1}$ for the HH~30 jet
and counterjet, and of $\sim$120~km~s$^{-1}$ for the HL~Tau jet.
Inclination angles with respect to the plane of the sky are
$0^\circ$--$40^\circ$ for the HH~30 jet and $60^\circ$ for the HL~Tau jet.
Comparison with previous observations suggests that most of the jet knots
consist of persisting structures. Also, we corroborate that the HH~30-N
knots correspond to the head of the HH~30 jet.  The overall HH~30 jet
structure can be well described by a wiggling ballistic jet, arising
either by the orbital motion of the jet source around a primary or by
precession of the jet axis because of the tidal effects of a companion. In
the first scenario, the orbital period would be 53~yr and the total mass
0.25-2~$M_\odot$. In the precession scenario, the mass of the jet source
would be $\sim0.1$--1~$M_\odot$, the orbital period $<1$~yr, and the mass
of the companion less than a few times $0.01~M_\odot$, thus being a
substellar object or a giant exoplanet. In both scenarios a binary system
with a separation $<18$~AU ($<0\rlap.''13$) is required. Since the radius
of the flared disk observed with the {\em HST} is $\sim$250~AU, we 
conclude that
this disk appears to be circumbinary rather than circumstellar, suggesting
that the search for the collimating agent of the HH~30 jet should be
carried out at much smaller scales.
\end{abstract}

\keywords{
ISM: Herbig-Haro objects ---
ISM: individual (HH 30, HH 266) ---
ISM: jets and outflows ---
stars: formation ---
stars: individual (HL Tau, XZ Tau)
}

\section{Introduction}

The region encompassing the Herbig-Haro (HH) object 30 and the stars
HL/XZ Tau, in the northeastern part of the L1551 dark cloud, lies at a
distance of 140 pc (e.g., Kenyon et al.\ 1994). It is particularly rich in
HH jets, being one of the regions where this phenomenon was first
identified (Mundt \& Fried 1983).

The HH~30 outflow is considered a prototypical jet/disk system. It presents a
clear jet/counterjet structure, which has been described, e.~g., by Mundt et
al.\ (1987, 1988) and by Graham \& Heyer (1990). The HH~30 exciting source is
an optically invisible star (Vrba, Rydgren, \& Zak 1985) highly extincted by an
edge-on disk (Burrows et al.\ 1996, Stapelfeldt et al. 1999), that extends up
to a radius of $\sim250$ AU perpendicularly to the jet, and divides the
surrounding reflection nebulosity into two lobes. Kenyon et al.\ (1998) propose
a spectral type M0 for the HH~30 star, and Cotera et al.\ (2001) estimate a
bolometric luminosity of 0.2--0.9 $L_\odot$. L\'opez et al.\ (1995, 1996) argue
that a number of knots located to the northeast of the HH~30 object are
possibly also part of the same flow, resulting in a total angular size of
$\sim7'$ for the whole outflow. Several studies have explored the spatial
morphology (both along and across the symmetry axis; Mundt et al.\ 1991, Ray et
al.\ 1996), line ratios (Mundt et al.\ 1990; Bacciotti, Eisl\"offel, \& Ray
1999), radial velocities (Raga et al.\ 1997), and proper motions (Mundt et al.\
1990; Burrows et al.\ 1996; L\'opez et al.\ 1996) of the HH~30 flow.

HL Tau, located $\sim1\rlap.'5$ north from the HH~30 source, has been one
of the most intensively studied T Tauri stars. Since the first proposal
that this star is associated with a nearly edge-on circumstellar disk
(Cohen 1983), numerous studies have been carried out in order to image the
proposed disk (e.g., Sargent \& Beckwith 1987, 1991; Lay et al.\ 1994;
Lay, Carlstrom, \& Hills 1997; Rodr\'\i guez et al.\ 1994; Mundy et al.\
1996; Wilner, Ho, \& Rodr\'\i guez 1996; Looney, Mundy \& Welch 2000),
although recent studies suggest that the flattened molecular structure
around HL Tau is part of a larger molecular shell-like structure (Welch et
al. 2000). This star has been proposed as the source of a molecular
outflow (Calvet, Cant\'o \& Rodr\'\i guez 1983; Torrelles et al.\ 1987;
Monin, Pudritz, \& Lazareff 1996). HL Tau is also the source of a
collimated jet/counterjet system, that has been extensively studied (e.g.,
Mundt et al.\ 1987, 1988, 1990, 1991, Rodr\'\i guez et al.\ 1994). L\'opez
et al. (1995, 1996), on the basis of geometrical alignment and proper
motion measurements, propose that HH 266, an extended structure $\sim4'$
to the northeast of HL Tau, might correspond to the head of the HL Tau
jet.

XZ Tau, located $\sim25''$ to the east of HL Tau, is a close binary system
composed of a T Tauri star and a cool companion separated by $0\rlap.''3$
(Haas, Leinert, \& Zinnecker 1990). XZ Tau is also the source of an
optical outflow, as revealed, e.g., by the studies of Mundt et al.\ (1988,
1990), and more recently by direct evidence of the expansion of the
nebular emission, moving away from XZ Tau, in the spectacular sequence of
{\em HST} images of Krist et al.\ (1999). XZ Tau has also been proposed as an
alternative driving source for the molecular high-velocity gas observed
towards the HL/XZ Tau complex (Torrelles et al.\ 1987), and is located at
the center of the ring- or shell-shaped molecular structure imaged by
Welch et al.\ (2000).

Despite the numerous studies carried out in this region, proper motions
have only been measured for a relatively small number of knots of the HH
jets present in this region, using images obtained under quite unequal
conditions (different filters, sensitivities or telescopes). Mundt et al.\
(1990) measured proper motions for one knot of the HH~30 jet, and five
additional knots near HL/XZ Tau by comparing a Gunn $r$ filter image with
a [SII]+H$\alpha$ image, separated by an interval of four years; Burrows
et al.\ (1996) measured proper motions for the small scale structure of
knots in the HH~30 jet/counterjet from two {\em HST} F675W images
(including [SII], H$\alpha$, and [OI] lines) separated by one year, but
because of differences in sensitivity between both images, proper motion
measurements were restricted to the region within $5''$ from the star;
finally, L\'opez et al.\ (1996) obtained proper motions for five knots of
the HH~30 jet, and for HH 266, from two [SII] images separated by about
one year, but obtained with different instrumentation.

In the present paper, we used sensitive CCD frames obtained at two epochs
with the same instrument, in order to make a detailed study of the proper
motions along the region. Our observations are complementary to those
carried out with the {\em HST} telescope, since the {\em HST} covers with
very high angular resolution the brightest region near (a few arcsec) the
exciting sources, while our observations cover a much more extended, low
brightness region, up to $\sim5'$ from the exciting sources. We measured
proper motions for a number of knots much larger than in previous studies,
in order to obtain the full kinematics of the region. Given the proximity
of the region ($D=140$ pc), a time span of one year between the two images
allowed us to measure the proper motions with enough accuracy, and made
easier the identification of the knots.  We compared our proper motion
results with those obtained in previous studies in order to obtain the
time evolution of the jet structures.

For the HH~30 jet, we have also carried out high resolution spectroscopy to
measure the radial velocity along the jet, in order to obtain the full
kinematics of this object.

The paper is organized as follows. The observations and procedures used to
measure the proper motions are described in \S\ 2. In \S\ 3 we present
the results we obtained from our proper motion and spectroscopic measurements
and we compare them with those of previous observations reported in the
literature. In \S\ 4 we discuss the origin of the HH~30 jet wiggling
structure in terms of orbital motions of the jet source and precession of the
jet axis; we also discuss the large-scale structure of the HH 30 jet. 
In \S\ 5 we give our conclusions.

\section{Observations and Data Analysis}

\subsection{CCD Imaging}

CCD images of the HL Tauri region, including the HH~30 jet and the HH~30-N
and HH 266 emission structures, were obtained at two different epochs
(1998 November 20 and 1999 November 8).  The observations were made with
the 2.5~m Nordic Optical Telescope (NOT) at the Observatorio del Roque de
los Muchachos (La Palma, Spain).  The same setup was used for the two
runs. The images were obtained with the Andalucia Faint Object
Spectrograph and Camera (ALFOSC), equipped with a Ford-Loral CCD with
$2048\times2048$ pixels with an image scale of $0\farcs188$~pixel$^{-1}$.  
A square filter, with a central wavelength $\lambda=6720$~\AA\ and
bandpass $\Delta\lambda=56$~\AA\ (that includes the red [SII] 6717,
6731~\AA\ lines) was used for these observations. On the NOT, the nominal
ALFOSC field is $6\rlap.'5\times6\rlap.'5$. However, because of the square
geometry of the [SII] filter used and the misalignment with the CCD axes,
the effective field sampled through the filter was only $\sim5'\times5'$.

 The images were processed with the standard tasks of the
IRAF\footnote{IRAF is distributed by the National Optical Astronomy
Observatories, which are operated by the Association of Universities for
Research in Astronomy, Inc., under cooperative agreement with the National
Science Foundation.} reduction package. Individual frames were recentered
using the reference positions of several field stars, in order to correct
for the misalignments among them.

The ``first epoch'' NOT image was obtained by combining five frames of
1800~s exposure each, in order to get a deep [SII] image of 2.5 h of
integration time.  Typical values of seeing for individual frames were
$1''$--$1\farcs2$, resulting in a seeing of $1''$ for the final ``first
epoch'' image.  The ``second epoch'' NOT image was obtained by combining
eight frames of 1800~s exposure each in order to get a final image with a
total integration time of 4 h.  For this second epoch, the typical values
of seeing for individual frames were $0\farcs7$--$1''$, resulting in a
seeing of $0\farcs7$ for the final ``second epoch'' image.

Images were not flux-calibrated. However, we have performed aperture
photometry of comon field stars on the images of the two epochs and we
estimate that the deepness of the two images differs by less than 0.5 mag
in the filter used.

\subsection{Measurement of Proper Motions
\label{smpm}}

To measure the proper motions, the first- and second-epoch [SII] CCD final
images were converted into a common reference system by using the
positions of eleven field stars plus the position of the HH~30 star.  The
$y$ axis is oriented at a position angle (P.A.) of $30\rlap.^{\circ}6$,
roughly coincident with the HH~30 jet axis. The GEOMAP and GEOTRAN tasks
of IRAF were used to perform a linear transformation with six free
parameters taking into account relative translation, rotations and
magnifications between frames. After the transformation, the average and
rms of the difference in position for the reference stars in the two
images was $-0.08\pm0.34$ pixels in the $x$ coordinate and $-0.18\pm0.21$
pixels in the $y$ coordinate.

In order to improve the signal-to-noise ratio of the diffuse knot
structures, the aligned images were convolved with Gaussian filters using
the GAUSS IRAF task. For the knots nearest to the HH~30 star (i.e., knots
A--D and counterjet), a Gaussian filter with a FWHM of $0\farcs36$ was
applied to both images.  For the rest of the knots the Gaussian filter
applied to the images had a FWHM of $0\farcs63$.

We defined boxes that included the emission of the individual
condensations in each epoch, we computed the two-dimensional
cross-correlation function of the emission within the boxes, and finally
we determined the proper motion through a parabolic fit to the peak of the
cross-correlation function (see the description of this method in Reipurth
et al.\ 1992, and L\'opez et al.\ 1996). The uncertainty in the position
of the correlation peak has been estimated through the scatter of the
correlation peak positions obtained from boxes differing from the nominal
one in 0 or $\pm2$ pixels ($0\farcs38$) in any of its four sides, making a
total $3^4=81$ different boxes for each knot. The error adopted has been
twice the rms deviation of position, for each coordinate, added
quadratically to the rms alignment error.

For the knots near the HH~30 source (knots A--D and counterjet), the
intensity gradient in the jet direction made unreliable the usage of the
correlation method. For this region, we integrated the emission across the
jet over a width of 11 pixels ($2\farcs1$) and identified the position of
each knot along the jet through a parabolic fit after baseline removal.\ A
similar procedure, but integrating the emission along the jet for a
typical width of 6 pixels ($1\farcs1$) at the position of each knot,
allowed us to measure the $x$ coordinate of the knots. For these knots we
estimated that the major source of error was the residual misalignment of
the two images, and the uncertainty adopted for the proper motions was
$\sqrt{2}$ times the rms alignment error for each coordinate. However, we 
cannot discard that residual contamination from continuum emission 
constitutes an additional source of uncertainty in our proper motion 
measurements of the knots nearest to the HH~30 source.

\subsection{Optical Spectroscopy}

Optical spectroscopy of the HH~30 jet and HH~30-N was acquired during 1998
December 11 and 12 using the red arm of the double-armed spectrograph ISIS
(Carter et al.\ 1994), and a Tektronics CCD detector of $1024\times1024$
pixels, at the Cassegrain focus of the 4.2~m William Herschel Telescope
(WHT) at the Observatorio del Roque de los Muchachos (La Palma, Spain).  
The high resolution grating R1200R (dispersion of 0.41~\AA~pixel$^{-1}$),
centered at 6600~\AA\ and covering a wavelength range of 420~\AA\ (that
includes the H$\alpha$ and [SII] 6717, 6731 \AA\ lines), was employed. The
effective spectral resolution achieved was 0.7~\AA\
($\sim32$~km~s$^{-1}$). The angular scale was $0\farcs36$~pixel$^{-1}$.

In order to obtain the spectrum of the HH~30 jet, the $3\farcm7$ long slit
was centered on the HH~30 star with a P.A. of $30^{\circ}$. One exposure
of 600 s of the HH~30 jet was obtained on 1998 December 12, with a slit
width of $1\farcs5$.
 Four exposures of 1800 s each, with a total integration time of 2~h, were
obtained on 1998 December 11 through HH~30-N.  The slit was centered on
the position of knot NE (L\'opez et al.\ 1996), at a P.A. of $30^{\circ}$,
covering the emission from knots NA to NH.

The data were reduced using the standard procedures for the long-slit
spectroscopy within the IRAF package.
 The spectra were not flux-calibrated. The line-of-sight velocity as a
function of position along the HH~30 jet and HH~30-N has been obtained by
fitting multiple Gaussians to the observed [SII] 6717, 6731~\AA\ and
H$\alpha$ emission line profiles, using the SPLOT task of IRAF. The
Gaussian profiles are described in terms of line center, which is
transformed into heliocentric velocity, and line width, given as the full
width at half maximum (FWHM).

\section{Results}

\subsection{Overall Description}

In Figure \ref{figcromo} we show the [SII] image of the overall region,
about $5'$ in size, corresponding to the first epoch. The field includes
the HH~30 source, near the southern edge of the image, which appears at
the vertex of a cone of nebulosity that apparently extends along several
tens of arcsec. The HH~30 jet crosses the image from southwest to
northeast, and its wiggling as it propagates away from its exciting source
is clearly visible. The image also includes the HH~30-N knots, near the
northern edge of the image, which have been proposed to also belong to the
HH~30 jet (L\'opez et al.\ 1995).  Unfortunately only the very first knots
of the HH~30 counterjet fall inside the field.

The bright stars HL and XZ Tau and their associated jets are also included in
the observed field. A large fraction of the HL Tau counterjet is visible in the
southwestern part of the image, while the HH~266 object, proposed to be
also associated with HL Tau (L\'opez et al.\ 1995) falls near the northeastern
corner of the image. The young stellar object LkH$\alpha$~358 is also visible
near the western edge of the image.

Faint knots, corresponding to the emission-line knots K1-K3 identified by
Mundt et al.\ (1988) (see their Fig.~1), can be seen in our images
$\sim30''$ to the southeast of XZ Tau.  Additional knots, extending
$\sim40''$ to the east of K1-K3 can also be seen in our images. These
knots are connected by diffuse emission that extends further away, forming
an apparent ``ring'' with a radius of $\sim1\rlap.'5$, surrounding the
darker region observed towards the center of the frame.  We detected these
knots and the diffuse emission in our images at the two epochs, but we did
not find evidence for systematic proper motions. Thus, this apparent ring
probably arises as a result of an increase in extinction towards the
center of the field, because of a foreground clump. It is interesting to
note that a molecular ring, centered near XZ Tau, has been mapped in
$^{13}$CO by Welch et al.\ (2000). The dark region at the center of our
image coincides with the eastern side of the molecular ring, suggesting
that this dark region corresponds to the increase of extinction because of
this foreground molecular cloud, with the diffuse optical emission
probably originating at the edges of this molecular structure. HL~Tau
falls close to the western side of the molecular ring and, as noted by
Welch et al.\ (2000), the large arc of scattered light that appears to
pass through HL~Tau, apparently corresponds to the western side of the
molecular structure.

\subsection{HH~30}

\subsubsection{Proper Motions}

In Table \ref{tabpm} and Figure \ref{fignot99} we give the proper motion 
results obtained for the HH~30 jet using the procedures described in \S 
\ref{smpm}. The proper motion velocity has been calculated assuming a 
distance of 140~pc. In the proper motion calculations, the $y$ axis is at 
a position angle of $30\rlap.\degr6$, in the approximate direction of the 
jet. We have set the zero point of our coordinate system on the bright 
knot closest to the HH 30 star, labeled as A0. We identified this knot 
with the knot 95-01N of the {\em HST} image of Burrows et al.\ (1996). 
These authors estimate from a fit of the {\em HST} image to a flared disk 
using scattered-light models that the position of the exciting source is 
shifted from 95-01N, corresponding to a position $y=-0\farcs51$.

The nomenclature used for the knots maintains the single letter names used
by Mundt et al.\ (1990) and L\'opez et al.\ (1995). However, the number
after the letter indicates only an order of distance from the HH~30 star
and no attempt has been made to be consistent with previous observations.
Thus, for instance, the group of knots B is roughly at the same distance
from the HH~30 source in the images of Mundt et al.\ (1990), L\'opez et
al.\ (1995), and in this paper; however, knots B1, B2, and B3 identified
in the present images do not have a clear correspondence with knots B1 and
B2 of Mundt et al.\ (1990) (see the discussion below).

The proper motion velocities obtained along the HH~30 jet range from 
$\sim100$ to $\sim300$~km~s$^{-1}$, in a direction close to the jet axis. 
Near the source the velocities appear to be higher, with velocities of 
$\sim200$ to $\sim300$~km~s$^{-1}$ for knots A to D (corresponding to 
distances from $3''$ to $20''$ from the source; first panel of Fig.\ 
\ref{fignot99}), except for the edges of the B condensation (knots B1 and 
B3), where the velocities are lower ($\sim150$~km~s$^{-1}$).  The velocity 
of the central knot (B2) is similar to the other knots in the A--D region.  
The velocity decreases beyond knot E1 (at $35''$, with a velocity of 
$\sim200$~km~s$^{-1}$), reaching values of $\sim150$~km~s$^{-1}$ for knots 
E2 to E4 (from $40''$ to $50''$), and $\sim100$~km~s$^{-1}$ for knots G 
(at $75''$). Finally, the velocity increases slightly up to 
$\sim150$~km~s$^{-1}$ for knots H to I (from $90''$ to $135''$). It should 
be noted that up to knot E3, the direction of the velocity is in general 
quite close to the jet axis ($|\Delta\mathrm{P.A.}|\la10\arcdeg$), while 
beyond knot E3 some of the knots present a significant velocity component 
westwards, perpendicularly to the jet axis 
($|\Delta\mathrm{P.A.}|\simeq30\arcdeg$--$40\arcdeg$), so that the 
decrease in velocity is still more noticeable for the velocity component 
along the jet axis.

We have only been able to measure the proper motions of two knots of the
counterjet. For knot Z2 (at $4.5''$ from the source) the velocity is
$\sim250$~km~s$^{-1}$, similar to that of knot A2 in the jet. However we
measured a significantly lower velocity, of $\sim70$~km~s$^{-1}$, for knot Z1
(at $2.5''$). 
 We think that this abnormally low value of the velocity, as compared with
the values measured for the remaining knots of the jet and the counterjet,
is likely caused by contamination by scattered light and might not
represent a measure of a true acceleration in the counterjet.

The proper motions measured for the HH~30-N knots are on the average
aligned with the direction of the HH~30 jet. However, the values obtained
both for the velocity and the position angle, show a dispersion
considerably larger than in the knots of the HH~30 jet (see last panel in
Fig.\ \ref{fignot99} and Table \ref{tabpm}). This larger dispersion in the
measured values can be due to the fact that in the HH~30-N knots the
emission is fainter than in the HH~30 jet, but could also be a consequence
of the interaction between the head of the jet and its surroundings. The
global proper motion velocity of the HH~30-N structure is
$\sim120$~km~s$^{-1}$ with $\mathrm{P.A.}\simeq30\arcdeg$, similar to the
direction of the HH~30 jet, thus supporting the hypothesis that this group
of knots corresponds to the head of the HH~30 jet (L\'opez et al.\ 1996).
The velocity values for individual knots range from $\sim50$ to
$\sim300$~km~s$^{-1}$. The largest velocity is measured for knot NC, but
most of the knots have velocity values of $\sim50$~km~s$^{-1}$. It should
be noted that NF, that was identified as a HH knot in previous works,
appears very circular and compact in our higher quality images presented
here (see Fig.\ \ref{fignot99}); since the proper motion is compatible
with zero ($v_t=34\pm30$~km~s$^{-1}$), we conclude that it is
most probably a field star.

\subsubsubsection{Comparison with Previous Observations}

L\'opez et al.\ (1996) measure proper motions for knots C/D and E of the
HH~30 jet. The directions of the proper motions we measured are roughly
consistent with those derived by L\'opez et al.\ (1996). For the HH~30 jet
we obtained a similar velocity for knot E. For knot C/D where L\'opez et
al.\ (1996) note that they obtain an absurdly large velocity of
$\sim700$~km~s$^{-1}$, we now obtain a more reasonable value of
$\sim300$~km~s$^{-1}$. Mundt et al.\ (1990) report a still lower proper
motion velocity of $\sim150$~km~s$^{-1}$ for knot C (the only knot for
which they measure the proper motion), but since they use images of two
epochs obtained through different filters, we think that our value is more
reliable. Burrows et al.\ (1996), from {\em HST} observations, measure proper
motion velocities of $\sim100-250$~km~s$^{-1}$ in the inner $\sim5''$ of
the jet and velocities of $\sim250$~km~s$^{-1}$ for distances of $\la1''$
in the counterjet. These values are roughly in agreement with the
velocities we measured for knots A2, A3 in the jet and Z2 in the
counterjet.

Regarding HH~30-N, the only previous available proper motion measurements
are those of L\'opez et al.\ (1996) for knots ND, NE, and NF. Our results
confirm that the direction of the proper motion velocities of HH~30-N is
consistent with HH~30-N belonging to the HH~30 jet, thus supporting the
claim of L\'opez et al.\ (1996). However, we obtained significantly lower
values for the velocities. In particular, we conclude that knot NF is a
field star (see discussion above). Since the knots of HH~30-N are weak and
diffuse, we attribute these discrepancies to the lower sensitivity and
angular resolution of the images used by L\'opez et al.\ (1996).

In order to better illustrate the comparison with previous observations of 
the positions and proper motions of the HH~30 jet knots, we plotted in 
Figure \ref{figcpm} their positions as a function of time. The vertical 
axis is the distance along the jet, at a position angle of 
$30\rlap.\degr6$. The five vertical lines of each panel mark the epoch of 
different observations and the circles indicate the position of the knots. 
Knots are labeled with the last two digits of the year of the observation, 
followed by the name used by each author. The vertical line with knots 
labeled ``87'' corresponds to the observation carried out in 1987 January 
by Mundt et al.\ (1990). Knots labeled ``93'' correspond to the 
observation carried out in 1993 December by L\'opez et al.\ (1995). Knots 
labeled ``95'' were observed with the Hubble Space Telescope in 1995 
January by Burrows et al.\ (1996). {\em HST} observations carried out in 
1994 February and 1995 March reported in Burrows et al.\ (1996) and Ray et 
al.\ (1996) have not been included for clarity of the figure. The two 
vertical lines labeled ``98'' and ``99'' correspond to the present 
observations with the NOT telescope. The dotted lines indicate the proper 
motions of the knots, as measured from the 1998 and 1999 observations, 
taken from Table~\ref{tabpm}. The shaded area along the proper motion line 
of knot 99-C indicates the formal uncertainty of the proper motion 
measurement for this knot, as it propagates with time. The uncertainties 
for other knots are similar to this case, shown here as an example. We 
have not included the data of the observations described in L\'opez et 
al.\ (1996), observed in 1995 February since they are of much lower 
quality.

The extrapolation of the proper motions measured from our NOT 1998-1999 
data to the epoch of the previous observations intersects the vertical 
lines at points that, in general, are in agreement with the positions of 
observed knots. However, this correspondence is better for knots located 
farther away from the exciting source, while for knots closer to the 
source the correspondence is more complex. As the right panel of Figure 
\ref{figcpm} illustrates, a backward extrapolation of our proper motion 
measurements for knots E1-E4 and G1-G2 (at distances of $\sim30''$ to 
$\sim70''$ from the source) provides a good agreement with the positions 
of the knots E and G actually observed in both the 1987 and 1993 images. 
The extrapolated position of knot 99-D1 falls very close to the position 
of knot 95-14N in the {\em HST} image, while the extrapolated position of 
knot 99-D2 falls outside the region where the {\em HST} observations are 
sensitive enough to detect the jet emission. In the 1993 image, the 
extrapolation of the proper motions of knots 99-D1 and 99-D2 falls in a 
region where it is difficult to separate the emission into several knots, 
and it is designated as 93-D1, while the D2 knot in the 1993 image would 
probably be displaced too far away (and, thus, too faint) in the 1998-1999 
images to be detected. Interestingly, our proper motion extrapolation 
suggests that knots D (at $\sim18''$ from the source) in the 1998-1999 
images arise from knot C (at $\sim14''$ from the source) in the 1987 
image, while knots D in that image would have faded away in our 1998-1999 
images. Knot C in our 1998-99 images appears to correspond to knot 93-C, 
and perhaps to knot 95-13N. For knots B1--B3 in the 1998-1999 images, the 
extrapolation suggests a correspondence with knots 11N and 12N in the 1995 
image, and knots B1 and B2 in the 1987 image. As we noted previously, we 
derived abnormally low values for the proper motions of knots 99-B1 and 
99-B3; an extrapolation of their positions using the proper motion value 
obtained for B2 would result in a one-to-one correspondence between knots 
99-B1, 99-B2 and 99-B3 with knots 95-10N, 95-11N, and 95-12N, 
respectively. Also, within current uncertainties, knot 87-B2 could 
correspond either to the extrapolation of knot 99-C or knot 99-B3.

For knots within $5''$ from the source, the {\em HST} image provides much 
more detail than our ground based 1998-99 images, thus making less useful 
the backward extrapolation of the NOT proper motions, since each 98 and 99 
knot corresponds to more than one 95 knot. However, from the inspection of 
Figure \ref{figcpm}, and using the proper motions reported in Burrows et 
al.\ (1996), it is clear that at least some of the 1995 {\em HST} knots do 
appear to correspond to structures observed in our ground based images.
 For example, both the proper motion velocity of 170~km~s$^{-1}$ reported
by Burrows et al.\ (1996) for knot 95-06N\,+\,95-07N, and our proper
motion velocity of $213\pm43$~km~s$^{-1}$ obtained for knot 99-A3,
suggests that their extrapolated positions coincide. Also the proper
motion velocity of 260~km~s$^{-1}$ reported for knot 95-02N, leads its
extrapolated position to coincide with that of knot 99-A2, for which we
obtained the same value of the proper motion velocity. Knot 99-A1, which
only appears in our 1999 NOT image, had probably still not emerged at the
epoch of the {\em HST} observations.  The extrapolated position for the
counterjet knot 95-02S with a proper motion velocity of 280~km~s$^{-1}$
reported by Burrows et al.\ (1996) appears to correspond with the position
of knot 99-Z2, while it is not clear if knot 95-01S corresponds to knot
99-Z1, given its proximity to the source that makes difficult its proper
identification in our NOT images.

Globally, it seems that near the source the interaction of the knots with the
medium is stronger, and fading out is more important, and thus the changes in
intensity and shape of structure makes more difficult the cross-identifications
of knots in different epochs.

\subsubsection{Radial Velocity\label{rvel}}

Since close to the HH~30 star the [SII] emission of the jet is stronger
than the H$\alpha$ emission, being also less contaminated by the emission
of the star, the heliocentric radial velocities along the HH~30 jet have
been derived from the [SII] line profiles. The [SII] 6717, 6731~\AA\
spectra were averaged over regions comparable to the size of the knots and
the heliocentric radial velocities obtained for each averaged region (from
knot F3 to I1) are listed in Table \ref{tabrjet} and shown in Figure
\ref{figdens}a. The lines are unresolved with our spectral resolution of
32~km~s$^{-1}$.
 The heliocentric velocity for the region within $40''$ of the source 
(knots F3 to E2+E3) remains almost constant, with a value of 
$\sim16$~km~s$^{-1}$, similar to that of the surrounding cloud 
(19~km~s$^{-1}$; see Mundt et al.\ 1990).  The heliocentric velocity 
increases with distance from 16~km~s$^{-1}$ at $40''$ (knots E2+E3) to 
47~km~s$^{-1}$ at $125''$ from the source (knot I1), defining a radial 
velocity gradient of $\sim0.4$~km~s$^{-1}$~arcsec$^{-1}$ between knots G 
and I1. These values are consistent with the preliminary results of Raga 
et al.\ (1997), although for the knots close to the source our velocities 
are $\sim10$--15~km~s$^{-1}$ lower. Using a value $v_t=200$ km s$^{-1}$ 
for the proper motions, we derive that the inclination angle, $\phi$, of 
the jet with respect to the plane of the sky ranges from $4^\circ$ to 
$9^\circ$ for distances to the source from $70''$ to $120''$. At smaller 
distances from the source, the jet lies essentially in the plane of the 
sky ($\phi\simeq0\arcdeg$). Burrows et al.\ (1996) and Wood et al.\ 
(1998), from a fit of the {\em HST} images to a flared disk using 
scattered-light models, derive a value of $\sim8\arcdeg$ for the 
inclination angle of the disk axis with respect to the plane of the sky. 
However, it should be noted that, although the values obtained for the 
inclination angle are similar, the derived axis of the jet points away 
from the observer, while the derived axis of the disk points towards the 
observer.

Since the [SII] lines are fainter than H$\alpha$ in HH~30-N (only knot NC
has been detected in [SII]), the radial velocities of HH~30-N have been
determined from the H$\alpha$ line profiles. After averaging the spectra
over regions comparable to the size of the knots, the heliocentric radial
velocity and the FWHM for each region were obtained (see Table
\ref{tabr30n}). The velocity dispersion in HH 30-N is much larger than in
the HH 30 jet, with values of the FWHM of the lines of typically
$\sim100$~km~s$^{-1}$. The values obtained for the heliocentric velocity
are blueshifted with respect to the ambient cloud, being in agreement with
our previous results (Raga et al.\ 1997). Using a value of $v_t\simeq120$
km s$^{-1}$ for the proper motions in HH30-N, we derive that the inclination
angle of the jet with respect to the plane of the sky is $\sim40\arcdeg$
(with the jet pointing towards the observer).

In order to study at higher spatial resolution the kinematical pattern we 
also performed Gaussian fits at different positions inside the knots of 
HH~30-N. The results are shown in Figure \ref{radial}, where the 
heliocentric radial velocity is plotted as a function of projected 
distance to the HH~30 star.  A more detailed inspection of 
Figure~\ref{radial} reveals that along knot NA the value of the 
heliocentric radial velocity progressively decreases (i.e., the absolute 
value of the velocity relative to the ambient cloud increases) from south 
to north, with values ranging from $\sim-70$~km~s$^{-1}$ to 
$\sim-90$~km~s$^{-1}$.  Towards knot NC there is no clear velocity 
gradient ($v_\mathrm{hel}\simeq-100$~km~s$^{-1}$), while the velocity 
increases from $\sim-100$~km~s$^{-1}$ to $\sim-50$~km~s$^{-1}$ from the 
southern edge of NE to the northern edge of NH. Globally, over the HH~30-N 
structure, the velocity increases from south to north with an overall 
gradient of $0.76$~km~s$^{-1}$~arcsec$^{-1}$.  The FWHM decreases slightly 
towards knot NH.

\subsubsection{Electron Density}

In addition to the kinematical information, our spectra also provide 
information on the electron densities, derived from the [SII] 6717, 6731 
\AA\ line ratio. Densities have been obtained assuming a temperature in 
the [SII] emitting zone of $10^4$~K. Table \ref{tabrjet} and Figure 
\ref{figdens}b show the electron densities in the knots of the HH~30 jet. 
A high electron density has been derived for the central region (within a 
few arcsec of the source), followed by a rapid decrease for the outer 
knots.

Previous determinations of the electron density in HH~30 (Mundt et al.\
1990; Bacciotti et al.\ 1999) traced only the inner region ($\sim10''$) of
the jet.  Our results cover a much more extended region of the jet, up to
$\sim120''$ from the source.  The overall behavior of the electron
density, strongly decreasing with distance to the source, agrees with the
results of Mundt et al.\ (1990) and Bacciotti et al.\ (1999).  These
authors, with higher angular resolution, measure a strong decrease of
electron density from $\sim10^4$~cm$^{-3}$ to $\sim10^3$~cm$^{-3}$ in the
inner $5''$ of the jet and counterjet; this region approximately
corresponds to the region binned for our measurement of
$3.7\times10^3$~cm$^{-3}$ at zero position offset. For the region of the
jet between $\sim5''$ and $\sim10''$ from the source, Mundt et al.\ (1990)
give a density of $\sim1000$~cm$^{-3}$ while Bacciotti et al.\ (1999)
measure a decrease from $\sim1100$~cm$^{-3}$ to $\sim370$~cm$^{-3}$, 
consistent with our value of 430~cm$^{-3}$ measured at $11''$
from the source. At distances larger than $\sim20''$ the density falls to
values below $\sim100$~cm$^{-3}$.

In the spectrum of the HH~30-N region the [SII] 6717, 6731~\AA\ emission
lines were only detected at knot NC, for which an electron density of
$\sim380$~cm$^{-3}$ was derived. This increase in the electron density may
be produced by the interaction of the jet with an ambient medium of
locally enhanced density.

\subsection{HL/XZ Tau and HH 266}

In Table \ref{tabpmhl} and in Figure~\ref{fighltau} we show the proper
motion results obtained for the knots near HL/XZ~Tau and for HH~266. The
nomenclature used for the knots near HL/XZ~Tau is consistent with that
used by Mundt et al.\ (1990). These authors consider that knots HL-A and
HL-E belong to the HL Tau jet, and HL-B to HL-G belong the HL Tau
counterjet. According to Mundt et al.\ (1990), knots A--D define another
jet emanating from a source, VLA 1, with a counterjet defined by knot J.
However, the existence of the source VLA 1 is highly doubtful, as shown by
more sensitive observations carried out by Rodr\'{\i}guez et al.\ (1994).
Our results for the proper motions show no significant differences in the
values of the velocity and position angle for knots HL-A, HL-E, A, B, and
D. Thus, our results are consistent with all these knots belonging to the
same jet, emanating from HL~Tau. Typical values are $\sim120$~km~s$^{-1}$
for the velocity and $\sim45\arcdeg$ for the P.A., similar to the position
angle of the jet axis. Proper motion measurements are more difficult for
the knots of the counterjet because many of them are weak and appear split
into several subcondensations; in fact, our results are the first
determination of proper motions in the HL~Tau counterjet. For all the
counterjet knots the proper motion velocities are pointing away from
HL~Tau. For those with a better determination of their proper motion, HL-C
and HL-G, the velocity is $\sim120$~km~s$^{-1}$, similar to the values
found for the jet knots, and the position angle is $\sim-110\arcdeg$. For
knot J we measure a very small proper motion, which does not support the
claim of Mundt et al.\ (1990) that it belongs to the VLA~1 counterjet.
Instead, the proper motion of knot J appears to point away from the
position of HL~Tau as the nearby H$\alpha$B--C knots do too. Perhaps this
group of knots could constitute a low velocity, poorly collimated ejection
from HL~Tau (at $\mathrm{P.A.}\simeq170\arcdeg$), although given their
very low velocities we do not discard that they constitute a nearly static
condensation.

For HH~266 we measured a proper motion velocity of $\sim100$~km~s$^{-1}$ 
with a P.A. of $43\arcdeg$.  L\'opez et al.\ (1996) note that HH 266 lies 
at a position that is aligned within only a few degrees with the direction 
of the HL Tau jet ($\mathrm{P.A.}=45\arcdeg$), suggesting that HH 266 
constitutes the head of the HL Tau jet. Our proper motion results give 
support to this suggestion, since both the position angle of the HH~266 
proper motion, which points away from HL Tau, as well as the value of the 
velocity that is similar to the velocities measured for the knots of the 
HL Tau jet, give support to this interpretation. However, the uncertainty 
in the P.A. of the proper motion is not small enough to discard that 
another object nearby to HL Tau (such as XZ Tau or even the HH~30 star) 
was the driving source for HH 266.

Mundt et al.\ (1990) measure proper motions for several knots. In Table
\ref{tabhlxz} we list the values obtained by these authors. As can be seen
in the table, the direction of the proper motion velocities we have
measured for the HL Tau jet is similar to that found by Mundt et al.\
(1990). However, the values obtained by these authors are significantly
higher than ours. We attribute this discrepancy to the fact that Mundt et
al.\ (1990) derive their proper motions using images obtained with
different filters; so, we think that our values are more reliable.  In
order to obtain an estimate of the proper motions with a longer time
baseline, we used the [SII] images published in Mundt et al.\ (1990) to
obtain 1987 epoch positions for several of the knots.  By comparing these
positions with our NOT 1999 positions we derived proper motions with a
time baseline of 12.9 years. The results are given in Table \ref{tabhlxz}.  
As can be seen in the table, the proper motions obtained in this way are
in better agreement with our NOT 1998-1999 measurements.

Since the radial velocity of the HL Tau jet with respect to the ambient
cloud is $\sim-200$ km s$^{-1}$ (Mundt et al.\ 1990), and proper motions
are $\sim120$ km s$^{-1}$, we infer that the total velocity of the HL Tau
jet is $\sim230$ km s$^{-1}$, being the angle of the jet with respect to
the plane of the sky of $\sim60\arcdeg$ (towards the observer).

The radial velocity of the counterjet measured by Mundt et al.\ (1990) is
$\sim100$ km s$^{-1}$, while the proper motions are similar to those of
the jet. This results in a significant difference of $\sim20\arcdeg$
between the inclination of the jet and the counterjet with respect to the
plane of the sky, at scales $\ga10''$. Pyo et al.\ (2006) also noticed a
slight asymmetry of the same sign in the velocities of the jet and the
counterjet, resulting in a difference of $\sim4\arcdeg$ between the
inclination of the jet and the counterjet, at scales of less than $2''$

\section{Discussion on the HH~30 Jet Structure}

\subsection{On the Origin of the Wiggling of the Jet
\label{swiggling}}

As noted before, the HH~30 jet presents a wiggling morphology that is
particularly evident in the group of knots E1--E4 (see Fig.\
\ref{fignot99}b). If the wiggling was produced as the result of a true
helical trajectory of the knots as they move away from the exciting source
(e.g., by a deflection, as the jet propagates, because of the encounter
with a set of high density clumps), one would expect to see changes in the
direction of their proper motions, following a velocity pattern tangent to
the helical trajectory (e.g., one would expect knot E1 to move to the
right of the jet axis, while knots E2 and E3 to move to the left). Despite
the uncertainties in the $x$ component of the velocities, we do not see
evidence for such a systematic pattern; rather, the direction of the
proper motion vectors appears to be quite close to that of the axis of the
jet. Furthermore, when comparing, for example, the E1--E4 structure in our
images with the previous images of Mundt et al.\ (1990) and L\'opez et
al.\ (1995), a displacement can be seen of the whole knot structure,
keeping the same morphology, while a change in this morphology would be
expected if the knots were following an helical trajectory. This is
further illustrated in Figure \ref{figcpm}, where it can be seen that the
group of knots E1--E4 moves essentially as a whole from one epoch to
another, maintaining the same relative distance between knots. Thus, we
conclude that the motion of the knots is essentially ballistic and that
the observed wiggling of the jet structure is most likely produced by
variations in the direction of the ejection at the origin of the jet.

\subsubsection{Orbital Motion of the Jet Source\label{orbital}}

We will test the possibility that the observed wiggling in the HH~30 jet
results mainly from the orbital motion of the jet source around a binary
companion. Following the formulation given by Masciadri \& Raga (2002), we
will consider a ballistic jet (i.e., a jet where the fluid parcels
preserve the velocity with which they are ejected) from a star in a
circular orbit, and we will further assume that the ejection velocity
(measured in a frame moving with the outflow source) is time independent
and parallel to the orbital axis.

Let $m_1$ be the mass of the jet source, $m_2$ the mass of the companion,
and $m=m_1+m_2$ the total mass of the system.  We will call $\mu$ the mass
of the companion relative to the total mass, so that 
 \begin{eqnarray}
m_1&=&(1-\mu)\,m, \nonumber\\
m_2&=&\mu\,m.
\label{m1m2}
\end{eqnarray}
Let $a$ be the binary separation (i.e., the radius of the relative
orbit). Therefore, the orbital radius of the jet source with respect to
the binary's center of mass (i.e., the radius of the jet source 
absolute orbit) is
\begin{equation}
r_o=\mu\,a,
\label{mua}
\end{equation}
and the orbital velocity of the jet source is given by
\begin{equation}
 v_o={2\pi{}r_o\over\tau_o},
\label{eqvo}
\end{equation}
where $\tau_o$ is the orbital period. 

We will use a Cartesian coordinate system $(x',y',z')$, where $(x',z')$ are in
the orbital plane, being the $x'$ axis the intersection of the orbital plane
with the plane of the sky. The $y'$ axis coincides with the orbital 
axis, at an angle $\phi$ with respect to the plane of the sky.  The ejection
velocity of the jet will have a component in the orbital plane due to the
orbital motion, $v_o$, and a component perpendicular to this plane,
$v_j$, assumed to be constant.  In this coordinate system the
shape of the jet is given by
\begin{equation}
\frac{x'}{r_o}= \kappa\frac{|y'|}{r_o}
\sin\left(\kappa\frac{|y'|}{r_o}-\psi\right)+
\cos\left(\kappa\frac{|y'|}{r_o}-\psi\right),
\label{eq8raga}
\end{equation}
where $\kappa\equiv v_o/v_j$, and $\psi$ is the orbital
phase angle (with respect to the $x'$ axis) at the epoch of observation.
The equation for the $z'$ coordinate is obtained by substituting $\psi$ by
$\psi+\pi/2$ in Eq.\ \ref{eq8raga}. Note that the jet ($y'>0$) and
counterjet ($y'<0$) shapes will have a reflection symmetry with respect to
the orbital plane.

If $D$ is the distance from the source to the observer,
the positions $(x,y)$ measured on the observed images (i.e., in the
plane of the sky) are given by
\begin{eqnarray}
x &=& \frac{x'}{D\,} , \nonumber\\
y &=& \frac{y'\cos\phi-z'\sin\phi}{D}\simeq \frac{y'\cos\phi}{D},
\end{eqnarray}
the last approximation being valid for a collimated jet with a small inclination
$\phi$ at distances large
enough from the source ($y'\gg z'$). The parameters of the model are directly
related to the observables: $\kappa$ is related to the half-opening angle,
$\alpha$, of the jet cone measured in the plane of the sky,
\begin{equation}
\kappa =\tan\alpha\,\cos\phi, 
\label{eqkappa}
\end{equation}
and the orbital radius $r_o$ is related to the observed 
period $\lambda_y$ of the wiggles (i.e.\ the angular distance in the plane of 
the sky between the positions of two successive maximum elongations),
\begin{equation}
r_o={\lambda_y\tan\alpha\over2\pi}D.
\label{eqro}
\end{equation}
In terms of these parameters, Eq.\ \ref{eq8raga} becomes
\begin{equation}
x=|y|\tan\alpha\,\sin\left[\frac{2\pi}{\lambda_y}(|y|-y_0)\right]+
\frac{\lambda_y}{2\pi}\tan\alpha\,
\cos\left[\frac{2\pi}{\lambda_y}(|y|-y_0)\right],
\end{equation}
where $y_0=\lambda_y\,\psi/2\pi$ is the offset from the origin, in the plane of
the sky, of the knot ejected when the source was at the $\psi=0$ position.

Wiggling in the HH~30 jet is most evident in the group of knots B, C, D, 
and E. Therefore, we used knots A1 to E4 to determine the parameters of 
the jet, and we found a very good match of the predicted shape with the 
observed image of the jet for 
$\alpha=1\rlap.\arcdeg43\pm0\rlap.\arcdeg12$, $\lambda_y=16''\pm1''$ 
(corresponding to $2240\pm140$~AU at a distance $D=140$ pc), and 
$y_0=4''\pm2''$ (corresponding to $560\pm280$ AU). The result is shown in 
Figure \ref{figwiggle}a. If this jet shape is translated to the epoch of 
the {\em HST} observations reported by Burrows et al.\ (1996) it is also 
consistent with the observed positions of the knots, thus confirming the 
validity of the fit.

Using Eq.~\ref{eqro} and the values obtained for $\alpha$ and $\lambda_y$
we derived an orbital radius $r_o=8.9\pm0.9$~AU (corresponding to
$0\rlap.''064\pm0\rlap.''006$). In addition, the fit allowed us to obtain
a more accurate value for the position angle of the axis of the jet,
$\mathrm{P.A.}=31\rlap.\arcdeg6$, in the range of offsets from the source
of $0''$ to $50''$.

Since the observed ratio of radial to proper motion velocities in the 
HH~30 jet is very small (see Tables \ref{tabpm} and \ref{tabrjet}), we 
infer that the inclination angle should be also very small, 
$\phi\la5\arcdeg$ (\S~\ref{rvel}). Then, from the proper motion 
measurements (Table 
\ref{tabpm}) we obtained an estimate of the ejection velocity, 
$v_j=v_t/\cos\phi\simeq200\pm50$~km~s$^{-1}$, and we 
derived the remaining orbital parameters. Using Eq.~\ref{eqkappa} we 
obtained the orbital velocity, $v_o=5.0\pm1.3$~km~s$^{-1}$, and 
using Eq.\ \ref{eqvo} we obtained the orbital period, 
$\tau_o=53\pm15$~yr. The line-of-sight component of the orbital 
velocity will produce an oscillation of the radial velocity along the path 
of the jet with a peak-to-peak amplitude of 
$2v_o\cos\phi\simeq10.0$~km~s$^{-1}$. This value is small 
compared to the spectral resolution of our observations ($\sim32$ 
km~s$^{-1}$), so it is not expected to produce significant oscillations in 
the observed radial velocity, consistent with what is observed (see 
Table~\ref{tabrjet} and Fig.~\ref{figdens}). In the case of the HH 43 jet, 
noticeable radial velocity oscillations with a peak-to-peak amplitude of 
16~km~s$^{-1}$ were observed with higher spectral resolution by Schwartz 
\& Greene (1999).

The total mass of the binary system is given by
\begin{equation}
\left(\frac{m}{M_\sun}\right)=
\mu^{-3}
\left(\frac{r_o}{\mathrm{AU}}\right)^3
\left(\frac{\tau_o}{\mathrm{yr}}\right)^{-2},
\label{mbin}
\end{equation}
 corresponding to $m=0.25\,\mu^{-3}\,M_\sun$, for the values of
$r_o$ and $\tau_o$ derived above. For a system with two
stars of the same mass ($\mu=0.5$), each component will have a mass of
$1.0\pm0.3~M_\sun$, and the separation between the two components would
be $a=r_o/\mu=17.8$~AU ($0\rlap.''128$). Values of
$\mu<0.5$ (corresponding to the jet source being the primary) would result
in $m>2~M_\sun$, and appear unlikely, given the estimated low bolometric
luminosity of the system ($L\simeq0.2$--$0.9\,L_\sun$ that, according to
D'Antona \& Mazzitelli 1997 would correspond to stellar masses roughly in
the range 0.1--1~$M_\sun$). Smaller masses would be obtained if $\mu>0.5$
(corresponding to the jet source being the secondary), with a lower limit
of $m_2=0.25~M_\sun$ ($\mu=1$), and a separation $a=r_o$. Thus,
under this scenario we expect that the exciting source of the HH~30 jet
belongs to a close binary system, with the two components separated by
$0\rlap.''064$--$0\rlap.''128$, and the total mass of the system in the
range 0.25--2~$M_\sun$.

\subsubsection{Precession of the Jet Axis}

An alternative possibility is that the observed wiggling of the HH~30 jet
is due to precession of the ejection axis of the jet, being driven by tidal
interactions between the disk from which the jet originates and a
companion star in a non coplanar orbit. For this model we will neglect the
orbital motion of the jet source, and we will assume that the wiggling of
the jet is the result of the changing direction of ejection of the jet.
Masciadri \& Raga (2002) show that the shape of the jet is given by
\begin{equation}
x'=y'\,\tan\beta\,
\cos\left(\frac{2\pi}{\tau_p}
\frac{|y'|}{v_j\cos\beta}-\psi\right),
\label{eq19raga}
\end{equation}
where $\beta$ is the angle between the central flow axis and the line of
maximum deviation of the flow from this axis, and $\tau_p$ is the
precession period.  Note that in this case the jet ($y'>0$) and counterjet
($y'<0$) shapes will have a point symmetry with respect to the jet source.

The precession angle, $\beta$, is related to the observables $\alpha$ (the 
half-opening angle of the jet cone measured in the plane of the sky) and 
$\phi$ (the inclination angle of the jet axis with respect to the plane of 
the sky) as,
 \begin{equation}
\tan \beta = \tan \alpha \cos \phi,
 \label{tanbeta}
 \end{equation}
 and the precesssion period $\tau_p$ is related to $\lambda_y$ (the
observed angular period of the wiggles) and $v_t$ (the measured proper
motion velocity, where $v_t=v_j \cos \phi$) as,
 \begin{equation}
\tau_p = \frac{\lambda_y D}{\tau_p v_t \cos \beta}.
 \label{taup}
 \end{equation}

Therefore, Eq.~\ref{eq19raga} can be rewritten in angular coordinates in 
the plane of the sky and in terms of observable parameters as,
 \begin{equation}
x=y\,\tan\alpha\,\cos\left[\frac{2\pi}{\lambda_y}(|y|-y_0)\right].
\end{equation}

An approximate expression relating the orbital and precession periods can
be derived from Eq.\ 24 of Terquem (1998), valid for a disk precessing as a
rigid body, by assuming that the disk surface density is uniform and that
the rotation is Keplerian,
\begin{equation}
\frac{\tau_o}{\tau_p}=
\frac{15}{32}\,\frac{\mu}{(1-\mu)^{1/2}}\,{\sigma^{3/2}\cos\beta},
\label{tauo}
\end{equation}
where $\mu=m_2/m$ is the ratio between the mass of the companion and
the total mass of the system, $\beta$ is the inclination of the orbital
plane with respect to the plane of the disk (coincident with half the
opening angle of the precession cone), and $\sigma=r_d/a$ is the
ratio of disk radius to binary separation.

 From the fit to the observed shape of the jet for knots A--E we obtained 
the values of $\alpha$ and $\lambda_y$ (\S~\ref{orbital}), and using 
Eqs.~\ref{tanbeta} and \ref{taup} we obtain a precession angle 
$\beta=1\rlap.\arcdeg42\pm0\rlap.\arcdeg12$ and a precession period 
$\tau_p=53\pm15$~yr, for $v_j=200$~km~s$^{-1}$ 
(\S~\ref{orbital}). The expected peak-to-peak oscillation of the observed 
radial velocity corresponding to this precession motion is 
$2\,v_j \sin\beta \cos\phi=9.9$~km~s$^{-1}$, a value similar to 
that expected in the case of pure orbital motion, and that is too small to 
be detectable, given the spectral resolution of our observations. Unlike 
the case of orbital motion, in the case of precession, the observables do 
not tightly constrain the orbital parameters, and a number of additional 
assumptions are required to infer their values. Since it is expected that 
the size of the disk is truncated by tidal interaction with the companion 
star in such a way that $1/4\le\sigma\le1/2$ (Lin \& Papaloizou 1993; 
Artymowicz \& Lubow 1994; Larwood et al.\ 1996; Terquem et al.\ 1999), we 
will adopt a value of $\sigma=1/3$, so that using Eq.\ \ref{tauo} the 
orbital period can be obtained from the observables as a function of only 
the parameter $\mu$. An additional constraint comes from the observed 
luminosity of the source that, according to D'Antona \& Mazzitelli (1997), 
suggests that the mass of the more massive of the two components should 
fall in the range 0.1--1~$M_\odot$. Finally, the hypothesis that the 
observed wiggling in the jet is mainly due to precession implies that the 
effect on the jet opening angle produced by the orbital motion of the jet 
source should be smaller than the precession angle 
($v_o/v_j<\tan\beta$).  According to Eq.\ \ref{eqkappa}, 
the orbital velocity of the jet source should be 
$v_o<5$~km~s$^{-1}$ in order to fulfill this condition.

Taking into account these constraints, we investigated the parameter space for
different values of $\mu$. For a given value of $\mu$ we calculated the orbital
period using Eq.~\ref{tauo}. Then, for each value of $v_o$ we derived
the value for $r_o$ by using Eq.~\ref{eqvo}, and using Eq.~\ref{mua}
the corresponding separation between the two stars. The mass of the binary
system was calculated using Eq.~\ref{mbin}, and the corresponding masses of the
two components were calculated using Eq.~\ref{m1m2}.
 Following this procedure we found that the orbital velocity and
luminosity constraints lead to the result that the more massive of the two
components (with a mass between 0.1 and 1~$M_\odot$) should be the jet
source.  The maximum orbital velocity of 5 km s$^{-1}$ gives the largest
values of the mass of the companion ($m_2=0.17~M_\odot$). Specifically,
for $v_o=5$~km~s$^{-1}$, the mass of the jet source should be in
the range $0.1<m_1<1~M_\odot$, resulting in a mass of the companion
$0.07<m_2<0.17~M_\odot$, a binary separation $1.04>a>0.86$ AU (corresponding to
an angular separation between
$0\rlap.''007$ and $0\rlap.''006$), and an orbital period
$2.5>\tau_o>0.8$ yr, respectively. The actual value of the
orbital velocity should be significantly lower than 5~km~s$^{-1}$,
resulting in very low values for the mass of the companion and the binary
separation. For instance, for $v_o=2$~km~s$^{-1}$, the mass of
the companion is $0.01<m_2<0.04~M_\odot$, the binary separation is
$a=0.33$ AU ($0\rlap.''002$), and the orbital period is
$0.6>\tau_o>0.2$ yr.  Therefore, in the precession scenario 
the companion is expected to be a brown dwarf star or even a giant
exoplanet.

\subsubsection{Evidence for a Binary Exciting Source}

After the discussion in the previous section we conclude that both the orbital
motion and the precession models are feasible. In the first case, the orbital
period would be 53 yr, the expected angular separation would be
$0\rlap.''064$--$0\rlap.''128$ (9--18 AU), and the jet source is expected to be
the secondary, while the mass of the primary is expected to fall in the range
0.25--1~$M_\odot$.   In the case of precession, in order to fulfill the
observational constraints, the jet source should be the primary, with a mass in
the range $\sim0.1$--1~$M_\odot$, resulting in much smaller values for the
derived parameters: the orbital period is expected to be less than $\sim1$ yr,
the mass of the companion less than a few times $\sim0.01~M_\sun$, and the
angular separation $<0\rlap.''007$ ($<1$ AU). 

We want to emphasize that both scenarios are consistent with the current 
observational data, and that both imply that the exciting source of the 
HH~30 jet should be a close binary (separation $<0\rlap.''1$). We take the 
very good agreement between the predicted and observed wiggling of the 
HH~30 jet as a strong (although indirect) evidence for the existence of 
such a binary system. A direct evidence would require to resolve the HH~30 
exciting source with an angular resolution better than $\sim0\rlap.''1$ in 
the case of orbital motions and better than $\sim0\rlap.''01$ in the case 
that the wiggling is originated by precession. Given the strong extinction 
towards the source and the high angular resolution required, observations 
at centimeter, millimeter, or submillimeter wavelengths will be necessary.  
In the first scenario, the required angular resolution can be currently 
achieved with the VLA although, unfortunately, the source appears to be 
weak at centimeter wavelengths, and has not been detected yet 
(Carrasco-Gonz\'alez et al. 2007).  In the precession scenario, the 
angular resolution required is near the limits of the expected 
capabilities of ALMA.

In both scenarios the range of values inferred for the mass of the system
is consistent with the estimate for the stellar mass of
$0.45\pm0.04~M_\sun$ obtained from IRAM Plateau de Bure $^{13}$CO
observations of the disk (Pety et al.\ 2006). We also note that
Stapelfeldt et al.\ (1999) find variability in the asymmetry of the disk
suggesting a characteristic time scale of 3 yr or less, which is of the
order of the values of the orbital period derived in the precession
scenario. This coincidence should be expected if the variability of the
illumination pattern were produced by an eclipsing binary system with this
orbital period.

The two proposed scenarios could be discriminated by taking into account
that in the first case mirror symmetry between the jet and counterjet is
expected, while in the second case point symmetry is expected to be found.
Unfortunately, our images do not cover the counterjet and we cannot
discriminate between both possibilities. We expect that future
observations will allow us to discriminate between the two scenarios.

We also note that, in both scenarios, the expected separation between the
two components of the binary system is $<18$ AU, a value much smaller than
the radius of the disk nearly perpendicular to the HH~30 jet observed with
the {\em HST} (Burrows et al.\ 1996, Stapelfeldt et al.\ 1999), which is
$\sim250$ AU. Given that the radius of any circumstellar disk associated
with the jet source should be smaller than the binary separation, this
implies that the {\em HST} disk should be a circumbinary disk and not a
circumstellar disk.  Also, since the scale of the jet collimating
mechanism should be much smaller than that of the mechanism that drives
the jet wiggling, which is of the order of the binary separation ($<18$
AU), we conclude that the $\sim$250 AU disk observed with the {\em HST} is
unlikely to have a relevant role in the jet collimation, contrary to what
has been thought up to now. These results suggest that the search for
the true collimating agent of the HH~30 jet (likely a circumstellar disk
associated with the jet source) should be done at very small angular
scales, $<0\rlap.''13$ ($<18$ AU).

\subsection{The Large-Scale Structure of the HH~30 Jet}

As a general trend, we observe that the direction of the proper motions
measured for the knots of the HH~30 jet approximately coincides with that
of the geometrical axis of the jet (see Fig.\ \ref{fignot99} and Table
\ref{tabpm}). There is, however, some indication of a systematic velocity
component perpendicular to the axis of the jet, so that the resulting
proper motion velocities deviate to the right of the jet axis (i.e.,
the velocities are orientated at a P.A. that, in general, is smaller than
that of the jet axis, whose P.A. is $\sim30^\circ$). The presence of this
velocity component perpendicular to the jet axis is in agreement with the
suggestion of L\'opez et al.\ (1995) that the observed ``axial rotation''
effect in the HH~30 jet (i.e., the direction of the axis of the jet and
counterjet curves westward as one moves away from the exciting source) is
a consequence of the relative motion between the source and the
environment. Such a scenario has been modeled in detail by Cant\'o \& Raga
(1995). The proximity of the powerful L1551-IRS5 molecular outflow, to the
southeast of the HH~30 jet, could also contribute to the velocity
component perpendicular to the jet axis.

\begin{figure}
\plotone{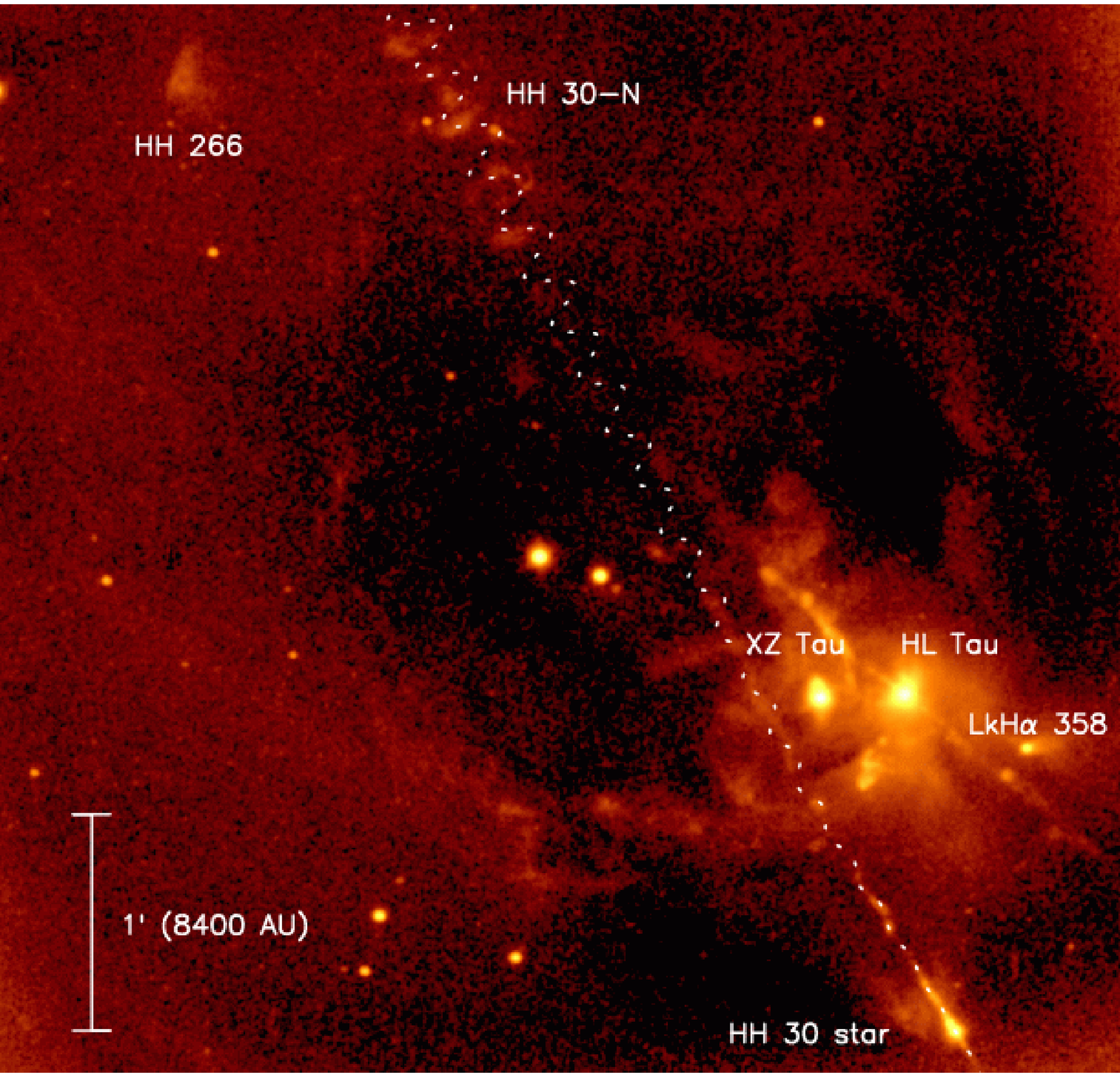}
\caption{
NOT 1998 [SII] image of the region surrounding the HH 30 jet, and the
HL/XZ Tau stars. North is up and east is to the left. The dotted
line is the shape of a wiggling jet model for the HH 30 jet (see 
text).
\label{figcromo}}
\end{figure}

\begin{figure}
\plotone{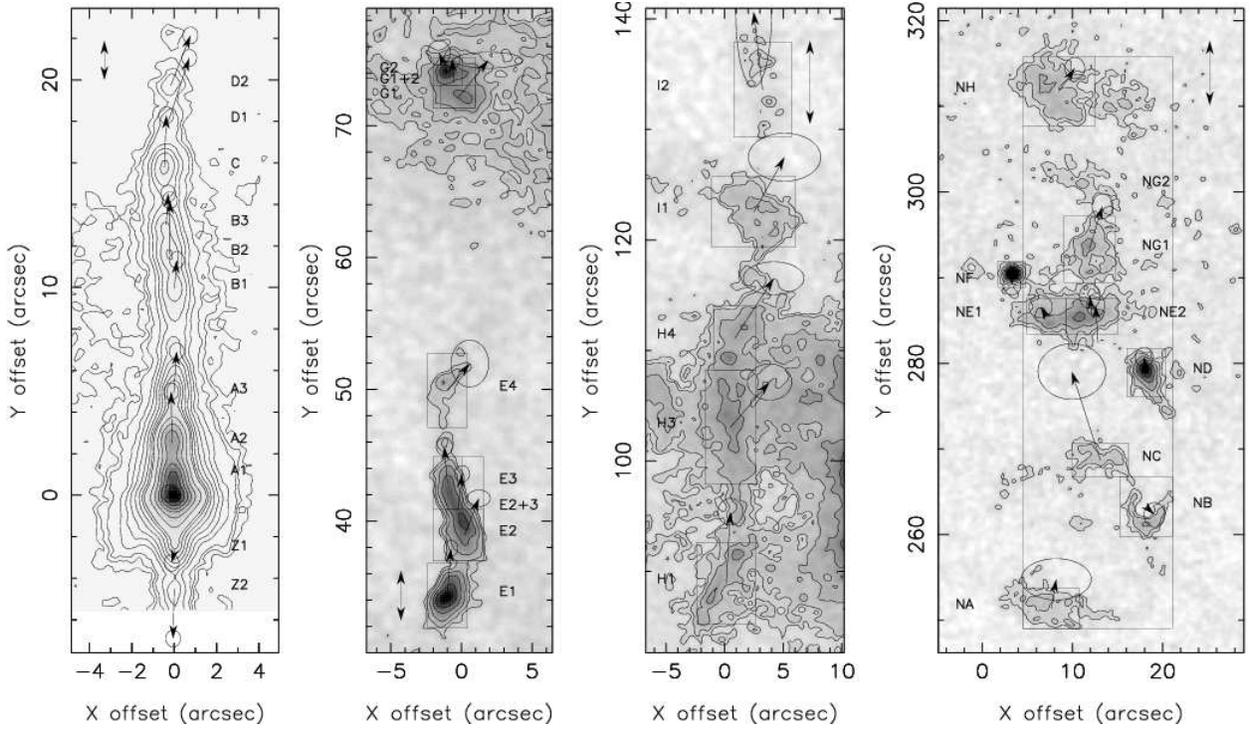}
\caption{ 
Proper motions of the HH~30 jet, shown on the NOT 1999 [SII] image. In {\em
(a)} we show the knots A to D and the counterjet, in {\em (b)} knots E to
G,  in {\em (c)} knots H to I, and in {\em (d)} knots N.  For knots E to N
we also show the box used for calculating the proper motion of each knot.
The $y$ axis, along the jet, is at a position angle of $30\rlap.\degr6$.
Both axes are labeled in arcsec, measured from the position of knot A0 (see
Table 1). 
Arrows indicate the proper motion velocity of each knot. The origin
of each arrow is at the peak position of the corresponding knot (A to D), or
at the center of the box used for calculating the proper motion (E to N).
Ellipses at the end of each arrow indicate the uncertainty in the components
of the velocity vector. The scale of the arrows is indicated by the double
headed arrow shown in a corner of each map, corresponding to a velocity of
200~km~s$^{-1}$.
\label{fignot99}}
\end{figure}

\begin{figure}
\plottwo{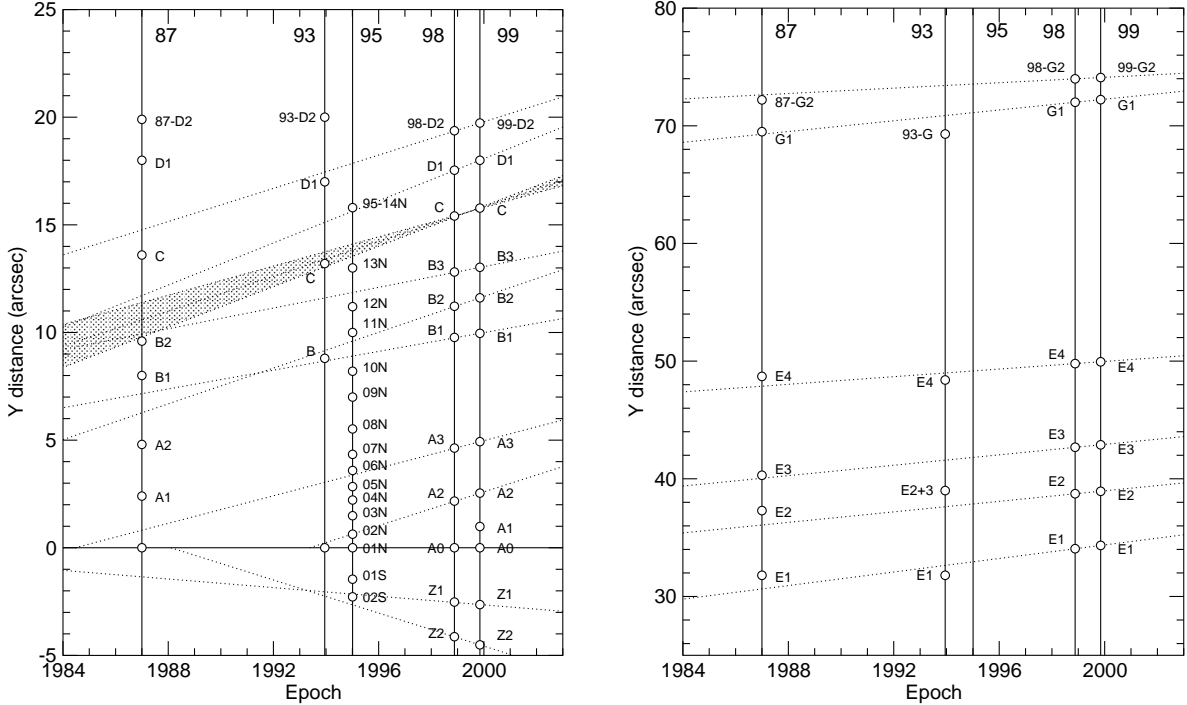}{f3right.eps}
\caption{
Positions of the knots in the HH~30 jet as a function of time. The vertical $y$
axis is in the direction of the jet axis. The zero of the $y$ axis has been set
to the position of the brightest knot in the ground-based observations, A0,
which we identified with knot 95-01N in the {\em HST} image.  Three 
vertical lines of
each panel mark the epoch of previous observations (87: Mundt et al.\ 1990; 93:
L\'opez et al.\ 1995; 95: Burrows et al.\ 1996). The two rightmost vertical
lines indicate the 1998 and 1999 observations with the NOT telescope (this
paper), with the dotted lines indicating the extrapolated proper motions of the
knots. The shaded area along the proper motion line of knot 99-C indicates the
typical uncertainty of the proper motion measurement as it propagates with time.
Note that the scale of the $y$ axis is different in the two panels.  
\label{figcpm}}
\end{figure}

\begin{figure}
\epsscale{0.8}
\plotone{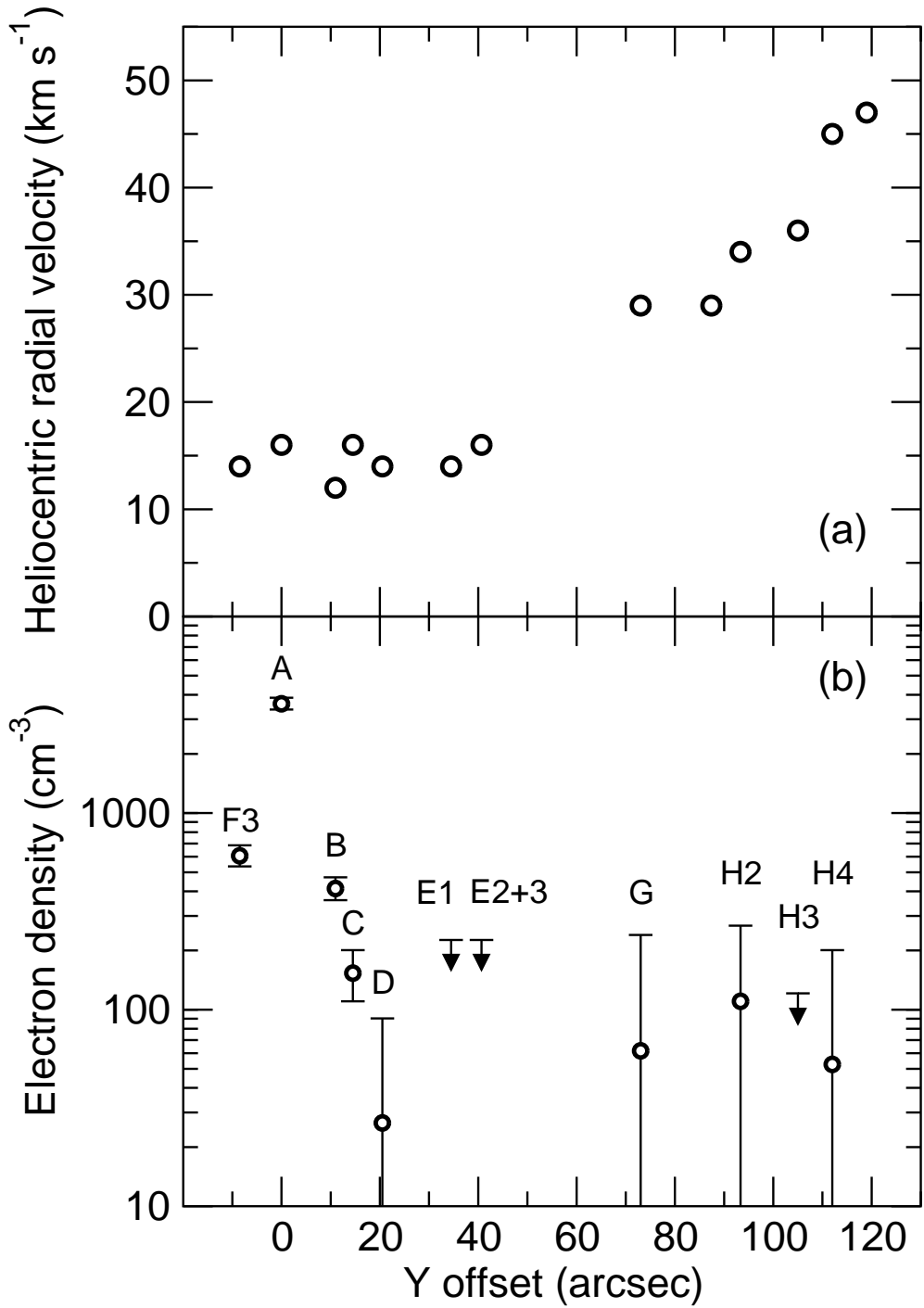}
\caption{
[SII] 6717 \AA\ heliocentric radial velocity \emph{(a)} and electron  density
\emph{(b)} of the HH~30 jet knots as a function of distance from knot A0
(see Table 1). The typical velocity uncertainty is $\sim$ 5 km 
s$^{-1}$. The heliocentric velocity of the ambient cloud is $\sim19$ km
s$^{-1}$ (Mundt et al. 1990). Error bars indicate the 
uncertainty in the derived electron density.
\label{figdens}}
\end{figure}

\begin{figure}
\epsscale{0.8}
\plotone{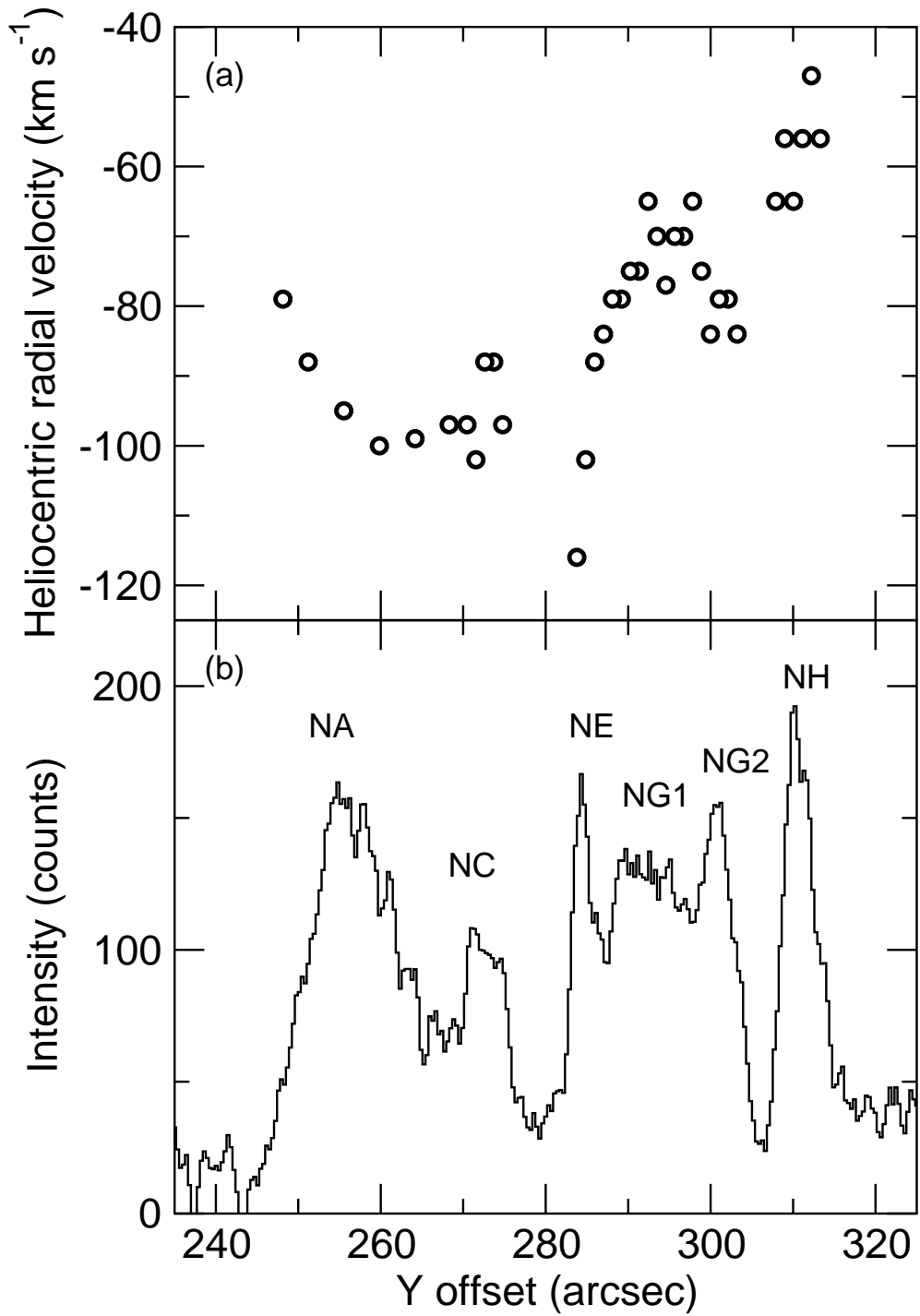}
\caption{
H$\alpha$ heliocentric radial velocities of HH~30-N as a function of distance
from knot A0 (see Table 1) (upper panel). The spatial profile along HH~30-N  of
the H$\alpha$ emission integrated over the width of the slit, is also  plotted
to aid in the location of the emission knots (lower panel). The typical 
velocity uncertainty is $\sim$ 5 km s$^{-1}$. The  heliocentric
velocity of the ambient cloud is $\sim$19 km s$^{-1}$ (Mundt  et al. 1990).
\label{radial}}
\end{figure}

\begin{figure}
\epsscale{0.55}
\plotone{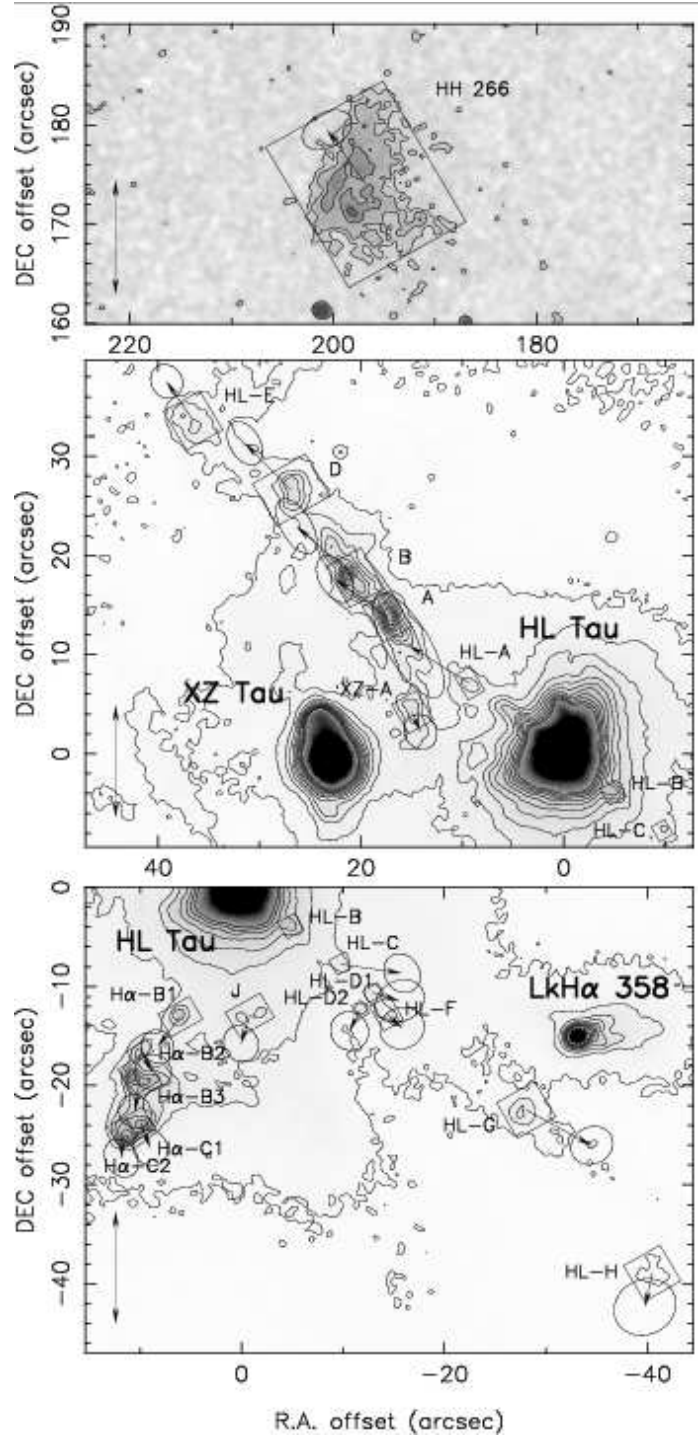}
\caption{\small{
Proper motions of the HL/XZ Tau region and HH 266, shown on the NOT 1999 [SII]
image. Lower panel shows the region south of HL Tau, central panel shows the HL
Tau and XZ Tau jets, and upper panel shows HH 266. Axes are labeled in right
ascension and declination offsets from the position of the HL Tau star. Arrows
indicate the proper motion velocity of each knot. Ellipses at the end of each
arrow indicate the uncertainty in the components of the velocity vector. The
scale of the arrows is indicated by the double headed arrow shown in the bottom
left corner of the map, corresponding to a velocity of 200~km~s$^{-1}$. 
\label{fighltau}}}
\end{figure}

\begin{figure}
\epsscale{0.8}
\plottwo{f7a.eps}{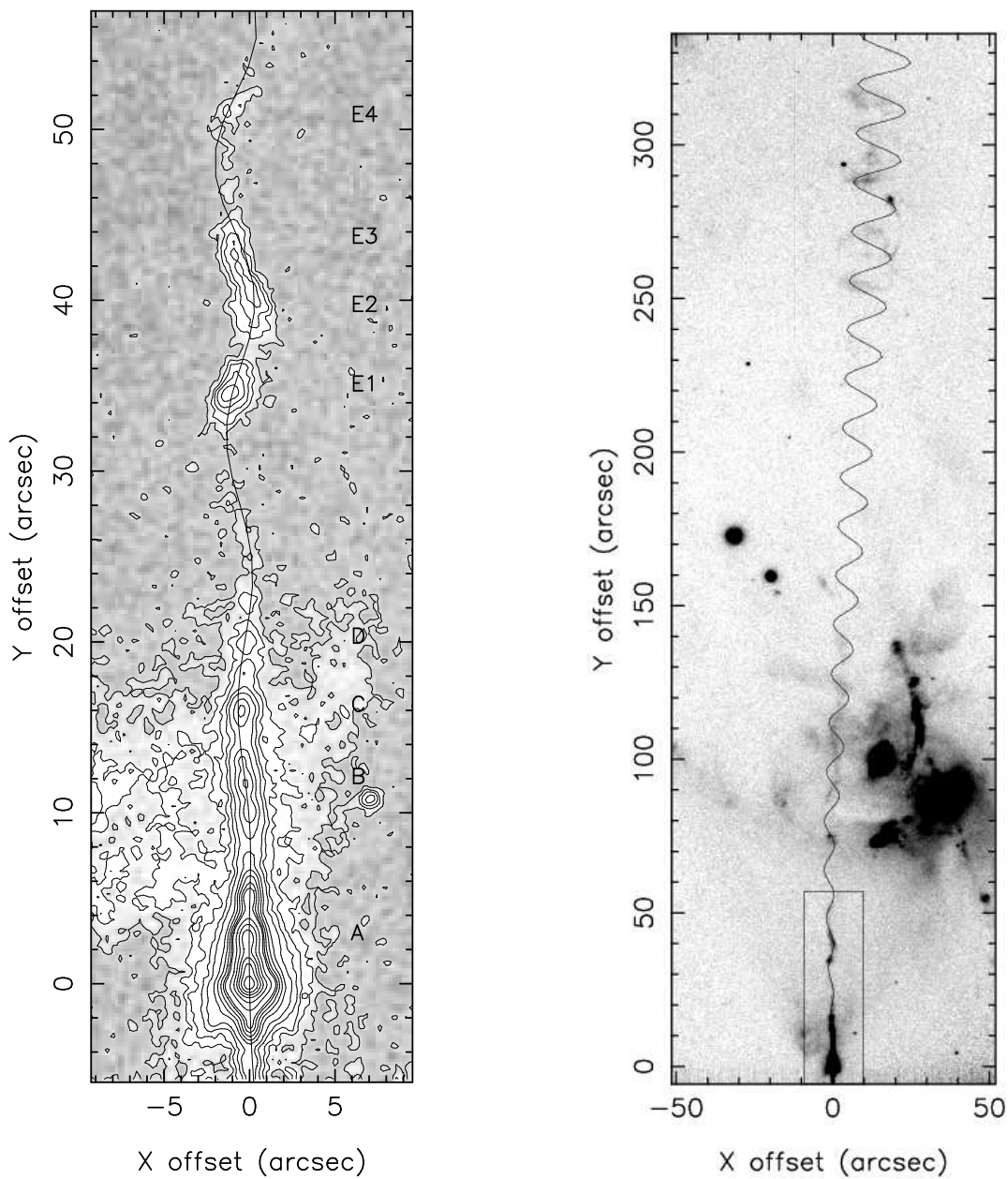}
\caption{
{\em (left)} NOT 1999 [SII] image of the A-E groups of knots of the HH 30 jet, 
with a fit of a ballistic jet shape overlaid. The parameters of the fit
are  $\mathrm{P.A.}=31\rlap.\degr6$, half-opening angle $\alpha=1\rlap.\degr43$,
separation between maxima $\lambda_y=16''$, and $y_0=4''$.  {\em (right)} Same
jet shape, but overlaid on the full image,  extending up to HH 30-N. The
P.A. changes in $-5\degr$ at a distance of $72''$ from the source.
\label{figwiggle}}
\end{figure}

The wiggling-model fit obtained for the jet at distances smaller than 
$\sim50''$ (see \S\ \ref{swiggling}) is also essentially valid for larger 
scales. In Figures \ref{figcromo} and \ref{figwiggle}b we show the 
resulting structure of the jet up to distances of $\sim300''$. In this fit 
we introduced a slight change of $-5^\circ$ in the P.A., at a distance of 
$72''$ from the source, in order to reproduce the bending of the jet (see 
discussion above). As can be seen in the Figure, in this way the overall 
structure of the jet is reproduced quite well, including the width of the 
jet up to the distance of HH~30-N. We take this as an additional proof 
that the HH~30-N knots do belong to the HH~30 jet (and that HH 266 does 
not belong to the HH~30 jet).  Also, we take this good agreement as 
additional evidence for the presence of a binary system in the exciting 
source of the HH~30 jet. 

We note that the detailed oscillations of the jet are not well reproduced 
for distances higher than $\sim50''$. This can be due to slight variations 
in the ejection velocity (in fact, proper motions are not constant) that 
would result in increasingly larger deviations of the periodic pattern as 
the distance from the source increases. In fact, a change in the radial 
velocity is observed at a distance of $\sim70''$ from the source, 
coinciding with the $\Delta\mathrm{P.A.}=-5^\circ$ of the jet.

Finally, we point out that if one wants to explain the knot/inter-knot 
pattern of the HH 30 jet as the result of source variability with a 
multiple mode model (as derived by Raga \& Noriega-Crespo 1998 for the 
HH~34 jet), we would need at least four modes. One of these modes would 
have a $\tau_1\simeq2.5$~yr period, as derived by Burrows et al.\ (1996) 
for the {\em HST} knots 02--04N. We would then need a second 
$\tau_2\simeq30$--40 yr period to explain the separation between features 
such as the knot that corresponds to the {\em HST} knots 06+07N and the 
condensation composed by the {\em HST} knots 02--04N (see Fig.\ 
\ref{figcpm}). A similar period is found from the velocities and 
separation between knots B and C, and knots E1 and E2 (see Figs.\ 
\ref{fignot99} and \ref{figcpm}).  A third, $\tau_3\simeq150$ yr period is 
necessary to produce the separations between knots B/C, E, G and H. 
Finally, in order to produce the NA-NH knot structure (see Figs.\ 
\ref{figcromo} and \ref{fignot99}), one would need a fourth source 
variability mode. The dynamical timescale 
$\tau_\mathrm{dyn}\simeq1500~\mathrm{yr}\simeq\tau_4$ of the NA--NH knots 
indicates the order of magnitude of the period of this fourth mode.

Even though there is little doubt that the {\em HST} knot spacing pattern is
generated very close to the outflow source (see Burrows et al.\ 1996), it
is clearly possible that the larger scale knot patterns could originate at
larger distances from the source (e.~g., through instabilities developed
along the jet beam). The question of whether or not these larger scale
knot patterns are also generated as a consequence of source variability
should be settled in the future through high angular resolution monitoring
of the HH~30 jet during the following $\sim30$ years.

\section{Conclusions}

Using the NOT we obtained [S~II] CCD frames at two epochs with a time span 
of one year of the region enclosing HH 30, HH 30-N, HH 266, as well as 
HL/XZ Tau. We also obtained high-resolution optical spectroscopy of the HH 
30 jet (including HH 30-N) using the WHT. The main conclusions from the 
analysis of our results can be summarized as follows:

\begin{enumerate}

\item
 We measured proper motions in the HH 30 jet, with velocities ranging from
$\sim100$ to $\sim300$ km~s$^{-1}$. We found the highest values of the velocity
(200--300 km~s$^{-1}$) near the driving source ($<3000$ AU), decreasing to
$\sim150$ km~s$^{-1}$ at distances of $\ga5000$ AU.

\item 
 Although the jet shows a wiggling morphology, the proper motions of the
jet knots are roughly parallel to the jet axis, with no signs of a pattern of
changes in the direction as would be expected for a true helical motion of the
knots. This suggests  that the motion of the knots is essentially ballistic and
that the observed wiggling is most likely produced by variations in the
direction of the velocity at the origin of the jet. Nevertheless, there is a
small but systematic drift of the velocities westwards, which could be due to
the effect of a side wind.

\item 
 We have been able to measure reliable proper motions only for one of the
knots of the counterjet, obtaining a velocity of $\sim250$ km~s$^{-1}$,
which is similar to the velocities measured in the jet at similar distances.

\item 
 The proper motions measured for the HH 30-N knots are, on the average,
aligned with the direction of the HH 30 jet, thus supporting the
hypothesis that this group of knots corresponds to the head of the HH~30
jet (L\'opez et al.\ 1996). However, the values obtained for both
magnitude and position angle of the velocity show a dispersion
considerably larger than in the HH 30 jet, which could be a consequence of
the interaction between the head of the jet and its surroundings.

Knot NF, which was previously identified as a HH knot, appears very circular
and compact in our higher quality images, and with a proper motion
compatible with being static. We thus conclude that it is most probably a
field star.

\item 
 The values we obtained for the proper motions along the HH 30 jet are similar
to those derived from {\em HST} data by Burrows et al.\ (1996) for the inner
($<5''$) region of the jet. We found a good agreement between the direction of
our proper motions and that obtained by L\'opez et al.\ (1996), although we
found discrepancies with the values of the velocity derived by these authors
(and by Mundt et al.\ 1990) that we attribute to the poorer quality of their
data.

\item 
 In general, we found a pretty good correspondence between the extrapolation
back in time of our proper motion estimates for the HH 30 knots and the
positions of the knots identified in previous observations. This result
indicates that most of the knots probably consist of persisting outflowing
structures. However, in some parts of the jet, particularly near the source
($\la20''$), where the interaction of the jet with the medium appears to be
stronger and fading is more noticeable than at larger distances, the knot
structure shows indications for an additional static pattern, that could arise
from the interaction with the ambient cloud.

\item 
 Our spectroscopic observations show that the radial velocity of the jet is
similar to the systemic velocity of the cloud.
 From the ratio of the radial to proper motion velocities we inferred that the
jet lies essentially in the plane of the sky ($\phi\simeq0^\circ$) for
distances to the source $<40''$. For distances to the source 
$70''<y<120''$ the jet is redshifted with respect to the ambient cloud, with an
inclination angle with respect to the plane of the sky $4^\circ<\phi<9^\circ$.
 For the HH 30-N structure, the radial velocity is blueshifted with respect to
the ambient cloud, with an inclination angle with respect to the plane of the
sky $\phi\simeq40^\circ$, suggesting a bending of the direction of the jet
propagation, as previously proposed by Raga et al.\ (1997).

\item 
 We estimated the electron density of the HH 30 jet knots up to 
$\sim120''$ from the source, and in HH 30-N, covering a region much more 
extended than in previous studies. The density in the HH 30 jet decreases 
with distance to the source, remaining below $\sim 100$~cm$^{-3}$ for 
distances $\sim20''$-$\sim120''$.  The density increases to 
$\sim400$~cm$^{-3}$ in HH 30-N, suggesting an interaction of the jet with 
an ambient medium of locally enhanced density.

\item 
 Our images reveal a clear wiggling of the HH 30 jet knots, with a spatial 
periodicity of $16''\pm1''$ ($2240\pm140$ AU). The width of the jet beam 
increases with distance to the source, with a half-opening angle in the 
plane of the sky of $1\rlap.^\circ43\pm0.^\circ12$. We found that the 
wiggling structure of the HH 30 can be accounted for either by the orbital 
motion of the jet source or by the precession of the jet axis.

In the first case the orbital period of the binary system would be 53 yr, 
the expected angular separation of the two components would be 
$0\rlap.''064$--$0\rlap.''128$ (9--18 AU), and the jet source is expected 
to be the secondary, while the mass of the primary is expected to fall in 
the range 0.25--1~$M_\odot$.  In the case of precession the jet source 
should be the primary, with a mass in the range $\sim0.1$--1~$M_\odot$. 
The orbital period should be less than 1 yr, the mass of the companion 
less than a few times $0.01~M_\odot$, and the angular separation 
$<0\rlap.''007$ ($<1$ AU). Therefore, it is feasible that the secondary is 
a substellar object, or even a giant exoplanet.

\item 
We take the very good agreement between the predicted and observed wiggling of
the HH~30 jet as a strong (although indirect) evidence for the existence of a
binary system. The angular separation between the two components is very small
($\la0\rlap.''1$ in the case of orbital motion and $\la0\rlap.''01$ in the case
of precession), and would require the use of VLA or ALMA to resolve the
system. 

In either case, the separation between the two components of the binary is
well below the size ($\sim450$ AU) of the observed disk perpendicular to
the jet, indicating that this disk should be a circumbinary disk instead
of a circumstellar disk, contrary to what has been thought up to now. This
leaves unclear the role of the observed disk in the jet collimation,
suggesting that the search for a circumstellar disk (likely the true
collimating agent of the HH~30 jet) should be carried out at very small
scales ($<0\rlap.''1$).

\item 
 Our fit of the observed knot structure of the jet allowed us to refine the 
value of the position angle of the axis of the jet, obtaining a value of
$\mathrm{P.A.}=31\rlap.^\circ6$. In fitting the large scale structure (up to
$300''$) we need to introduce a change in the direction of the jet,
$\Delta\mathrm{P.A.}=-5^\circ$ at a distance $\sim 70''$. This change in the
direction occurs roughly at the same position where there is a change 
in the radial velocities observed, suggesting a change in the
inclination angle with respect to the plane of the sky.

\item
We obtained a more accurate estimate of the proper motions of the HL Tau jet,
which are of the order of $\sim120$ km~s$^{-1}$, a value significantly lower
than previous estimates. From our proper motions and using the radial velocity
measurements of Mundt et al.\ (1990), we estimated that the inclination angle
of the HL Tau jet with respect to the plane of the sky is $\sim60^\circ$. We
measured for the first time the proper motions in the HL Tau counterjet,
obtaining values similar to those of the jet. 

\end{enumerate}

\acknowledgments

G. A., R. L., R. E., and A. R. are supported by the MEC AYA2005-05823-C03
grant (co-funded with FEDER funds). 
G.A. and J.M. acknowledge support from Junta de Andaluc\'{\i}a.
The work of A. C. R. was supported by the CONACyT.  
We thank Luis F. Miranda for his valuable comments and his help in 
preparing Figure 1.
We thank an anonymous referee for helpful comments.
The data presented here were taken using ALFOSC, which is owned by the
Instituto de Astrof\'{\i}sica de Andaluc\'{\i}a (IAA) and operated at the
Nordic Optical Telescope under agreement between the IAA and the NBIfA of the
Astronomical Observatory of Copenhagen.

\clearpage


\clearpage


\begin{deluxetable}{lrrrrr}
\small
\tablewidth{0pt}
\tablecaption{Proper Motions of Knots in the HH~30 Jet
\label{tabpm}}
\tablecolumns{8}
\tablehead{
&
\colhead{$y$\tablenotemark{a}}&  
\colhead{$\mu_x$}&
\colhead{$\mu_y$}&
\colhead{$v_t$\tablenotemark{b}}&
\colhead{P.A.\tablenotemark{c}}
\\
\colhead{Knot}&
\colhead{(arcsec)}& 
\colhead{(arcsec yr$^{-1}$)}&
\colhead{(arcsec yr$^{-1}$)}&
\colhead{(km s$^{-1}$)}&
\colhead{(deg)}
}
\startdata
Z2  & $ -4.5$& $-0.01\pm0.06$& $-0.38\pm0.06$& $255\pm\phn43$& $ 209\pm\phn8$\\
Z1  & $ -2.6$& $-0.02\pm0.06$& $-0.10\pm0.06$& $ 68\pm\phn43$& $ 196\pm   31$\\
A0  & $  0.0$& $-0.04\pm0.06$& $ 0.01\pm0.06$& $ 30\pm\phn38$& $ 101\pm   81$\\
A1\tablenotemark{d}
    & $  1.0$& \nodata       & \nodata       & \nodata       & \nodata       \\
A2  & $  2.5$& $-0.01\pm0.06$& $ 0.39\pm0.06$& $260\pm\phn43$& $  31\pm\phn8$\\
A3  & $  4.9$& $ 0.02\pm0.06$& $ 0.32\pm0.06$& $213\pm\phn43$& $  26\pm   10$\\
B1  & $ 10.0$& $ 0.02\pm0.06$& $ 0.22\pm0.06$& $145\pm\phn43$& $  25\pm   15$\\
B2  & $ 11.6$& $ 0.00\pm0.06$& $ 0.42\pm0.06$& $276\pm\phn43$& $  29\pm\phn8$\\
B3  & $ 13.0$& $ 0.01\pm0.06$& $ 0.24\pm0.06$& $160\pm\phn43$& $  27\pm   13$\\
C   & $ 15.8$& $ 0.02\pm0.06$& $ 0.40\pm0.06$& $268\pm\phn43$& $  28\pm\phn8$\\
D1  & $ 18.0$& $ 0.16\pm0.06$& $ 0.49\pm0.06$& $341\pm\phn43$& $  12\pm\phn6$\\
D2  & $ 19.7$& $ 0.15\pm0.06$& $ 0.39\pm0.06$& $275\pm\phn42$& $   9\pm\phn8$\\
E1  & $ 34.3$& $ 0.02\pm0.04$& $ 0.29\pm0.05$& $191\pm\phn33$& $  26\pm\phn8$\\
E2  & $ 38.9$& $ 0.08\pm0.06$& $ 0.22\pm0.05$& $158\pm\phn35$& $  10\pm   15$\\
E2+3& $ 40.9$& $ 0.02\pm0.04$& $ 0.21\pm0.05$& $143\pm\phn33$& $  24\pm   11$\\
E3  & $ 42.9$& $-0.02\pm0.05$& $ 0.22\pm0.05$& $147\pm\phn36$& $  35\pm   12$\\
E4  & $ 49.9$& $ 0.13\pm0.11$& $ 0.16\pm0.14$& $136\pm\phn86$& $  -8\pm   34$\\
G1  & $ 72.2$& $ 0.17\pm0.20$& $ 0.23\pm0.05$& $190\pm\phn82$& $  -7\pm   32$\\
G1+2& $ 73.3$& $-0.01\pm0.05$& $ 0.15\pm0.06$& $ 98\pm\phn38$& $  35\pm   18$\\
G2  & $ 74.1$& $-0.05\pm0.06$& $ 0.12\pm0.08$& $ 83\pm\phn50$& $  53\pm   30$\\
H1  & $ 89.0$& $ 0.01\pm0.04$& $ 0.26\pm0.05$& $172\pm\phn32$& $  27\pm   10$\\
H3  & $103.1$& $ 0.13\pm0.08$& $ 0.16\pm0.07$& $140\pm\phn46$& $  -9\pm   19$\\
H4  & $111.1$& $ 0.13\pm0.10$& $ 0.22\pm0.06$& $168\pm\phn49$& $  -1\pm   21$\\
I1  & $122.5$& $ 0.11\pm0.12$& $ 0.20\pm0.09$& $150\pm\phn64$& $   2\pm   30$\\
I2  & $133.6$& $-0.03\pm0.06$& $ 0.28\pm0.27$& $188\pm   180$& $  36\pm   14$\\
NA  & $251.4$& $ 0.02\pm0.16$& $ 0.14\pm0.10$& $ 92\pm\phn66$& $  21\pm   64$\\
NB  & $263.2$& $ 0.04\pm0.05$& $-0.04\pm0.06$& $ 34\pm\phn34$& $-105\pm   58$\\
NC  & $268.7$& $-0.13\pm0.15$& $ 0.42\pm0.13$& $288\pm\phn88$& $  47\pm   20$\\
ND  & $278.9$& $-0.01\pm0.04$& $ 0.07\pm0.05$& $ 48\pm\phn30$& $  37\pm   31$\\
NE1\tablenotemark{e}
    & $285.5$& $-0.02\pm0.05$& $ 0.04\pm0.05$& $ 34\pm\phn35$& $  59\pm   59$\\
NE2\tablenotemark{e}
    & $285.5$& $-0.01\pm0.05$& $ 0.09\pm0.06$& $ 62\pm\phn38$& $  36\pm   28$\\
NF\tablenotemark{f}
    & $290.4$& $ 0.00\pm0.04$& $-0.05\pm0.05$& $ 34\pm\phn30$& $-147\pm   44$\\
NG1 & $293.3$& $ 0.06\pm0.05$& $ 0.20\pm0.06$& $139\pm\phn38$& $  14\pm   13$\\
NH  & $311.7$& $ 0.07\pm0.04$& $ 0.11\pm0.05$& $ 86\pm\phn32$& $  -3\pm   20$\\
NA-H\tablenotemark{g}
    & $282.4$& $-0.01\pm0.04$& $ 0.18\pm0.05$& $117\pm\phn35$& $  33\pm   13$\\
\enddata

\tablenotetext{a}
{Angular distance from the position of the brightest knot, A0, at 
$\alpha$(J2000)=$04^\mathrm{h}31^\mathrm{m}37\fs450$, 
$\delta$(J2000)=$+18\degr12'25\farcs42$, measured on the 1999 [SII] image 
along a
direction with $\mbox{P.A.}=30\rlap.\degr6$, nearly coincident with the jet
axis. The HH 30 star is estimated to be at $y=-0\farcs51$ (Burrows et al.\
1996; see text). 
}
\tablenotetext{b}
{Proper motion velocity, assuming a distance of 140 pc.}
\tablenotetext{c}
{Position angle with respect to the north direction.}
\tablenotetext{d}
{Knot A1 has only been identified in the 1999 image. No proper motion
measurement is available.}
\tablenotetext{e}
{Knot NE has been split into two in order to minimize the effect of a clear
morphology change between the two epochs in the proper motion measurement.}
\tablenotetext{f}
{The image of knot NF is very circular and compact, and its proper motion is
compatible with zero. Most probably it is a field star.}
\tablenotetext{g}
{Box NA-H encompasses all the N knots, except the probable star NF.}

\end{deluxetable}


\begin{deluxetable}{lrrrrr}
\tablewidth{0pt}
\tablecaption{
Spectroscopy of the HH~30 Jet\tablenotemark{a}
\label{tabrjet}}
\tablehead{
&&&\multicolumn{2}{c}{$v_{\rm hel}$\tablenotemark{b}}\\
\cline{4-5}&
\colhead{$y$\tablenotemark{c}}&
\colhead{Size\tablenotemark{d}}&
\colhead{[SII] 6717}&
\colhead{[SII] 6731}&
\colhead{$n_e$}\\
\colhead{Knot}&
\colhead{(arcsec)}&
\colhead{(arcsec)}&
\colhead{(km s$^{-1}$)}&
\colhead{(km s$^{-1}$)}&
\colhead{(cm$^{-3}$)}
}
\startdata
F3    &$ -8.5$ & 4.4 & 14 & 16     & $ 630$ \\
A     &$  0.0$ &11.2 & 16 & 16     & $3670$ \\
B     &$ 11.0$ & 3.6 & 12 & 14     & $ 430$ \\
C     &$ 14.5$ & 4.4 & 16 & 14     & $ 174$ \\
D     &$ 20.5$ & 4.7 & 14 & 16     & $  40$ \\
E1    &$ 34.5$ & 3.3 & 14 & 20     & $< 10$ \\
E2+3  &$ 40.7$ & 7.6 & 16 & 16     & $< 10$ \\
G     &$ 73.0$ &12.6 & 29 & 29     & $  80$ \\
H1    &$ 87.4$ & 2.9 & 29 & 25     &\nodata \\
H2    &$ 93.4$ & 8.3 & 34 & 31     & $ 130$ \\
H3    &$105.0$ & 6.8 & 36 & 38     & $< 10$ \\
H4    &$112.0$ & 6.5 & 45 & 41     & $  70$ \\
I1    &$119.0$ & 3.2 & 47 &\nodata &\nodata   
\enddata

\tablenotetext{a}
{{Line} widths are smaller than the effective spectral  resolution
($\sim32$~km~s$^{-1}$).}
\tablenotetext{b}
{Error is $\sim5$~km~s$^{-1}$.}
\tablenotetext{c}
{Angular distance from knot A0 (see Table~\ref{tabpm}).}
\tablenotetext{d}
{Size of the region over which the spectra have been binned.}

\end{deluxetable}

\clearpage


\begin{deluxetable}{lrrrr}
\tablewidth{0pt}
\tablecaption{
Spectroscopy of HH~30-N\tablenotemark{a}
\label{tabr30n}}
\tablehead{
&
\colhead{$y$\tablenotemark{b}}&
\colhead{Size\tablenotemark{c}}&
\colhead{$v_{\rm hel}$\tablenotemark{d}}&
\colhead{FWHM}\\
\colhead{Knot}&
\colhead{(arcsec)}&
\colhead{(arcsec)}&
\colhead{(km s$^{-1}$)}&
\colhead{(km s$^{-1}$)}
}
\startdata
NA 		    & 258 &16.9 & $-96$ &103 \\
NC\tablenotemark{e} & 274 & 5.4 & $-89$ & 98 \\
NE 		    & 286 & 7.5 & $-91$ &103 \\
NG1		    & 293 & 7.9 & $-73$ &112 \\
NG2		    & 303 & 5.4 & $-73$ & 94 \\
NH 		    & 312 & 7.2 & $-50$ & 85
\enddata

\tablenotetext{a}
{Derived from the H$\alpha$ line.}
\tablenotetext{b}
{Angular distance from knot A0. H$\alpha$ spectroscopy data only provided
position offsets between knots. The values of $y$ have been
estimated by assigning to knot NE the angular distance from knot A0 
obtained from the 1999 [SII] image (Table~\ref{tabpm}).}
\tablenotetext{c}
{Size of the region over which the spectra have been binned.}
\tablenotetext{d}
{Error is smaller than $\sim10$~km~s$^{-1}$.}
\tablenotetext{e}
{The [SII] lines have been detected only for this knot, resulting in an
electron density $n_e=380$ cm$^{-3}$.}

\end{deluxetable}


\begin{deluxetable}{lrrrr}
\small
\tablewidth{0pt}
\tablecaption{Proper Motions of Knots in HL/XZ Tau and HH 266 
\label{tabpmhl}}
\tablecolumns{8}
\tablehead{
&
\colhead{$\Delta\alpha$\tablenotemark{a}}&  
\colhead{$\Delta\delta$\tablenotemark{a}}&  
\colhead{$v_t$\tablenotemark{b}}&
\colhead{P.A.\tablenotemark{c}}
\\
\colhead{Knot}&
\colhead{(arcsec)}& 
\colhead{(arcsec)}& 
\colhead{(km s$^{-1}$)}&
\colhead{(deg)}
}
\startdata
HL-E                 & $ 36.5$& $ 33.4$& $ 84\pm31$& $  34\pm19$\\
D                    & $ 26.8$& $ 26.4$& $117\pm42$& $  44\pm14$\\
B                    & $ 20.5$& $ 17.6$& $138\pm52$& $  48\pm13$\\
A                    & $ 17.3$& $ 13.7$& $114\pm43$& $  53\pm15$\\
HL-A                 & $  9.2$& $  6.9$& $126\pm95$& $  58\pm27$\\
HL-B\tablenotemark{d}& $ -4.9$& $ -3.9$& \nodata   & \nodata    \\
HL-C                 & $ -9.7$& $ -7.8$& $110\pm33$& $ -98\pm18$\\
HL-D1                & $-13.0$& $-10.6$& $ 49\pm40$& $-108\pm48$\\
HL-D2                & $-11.7$& $-12.4$& $ 42\pm33$& $ 154\pm46$\\
HL-F                 & $-14.3$& $-12.7$& $ 38\pm37$& $-134\pm59$\\
HL-G                 & $-27.9$& $-22.5$& $132\pm34$& $-119\pm16$\\
HL-H                 & $-40.5$& $-38.9$& $ 62\pm52$& $ 168\pm49$\\
J                    & $ -0.8$& $-12.9$& $ 54\pm32$& $ 163\pm34$\\
H$\alpha$-B1         & $  6.0$& $-13.0$& $ 67\pm28$& $ 144\pm26$\\
H$\alpha$-B2         & $ 10.0$& $-16.5$& $ 44\pm31$& $-150\pm42$\\
H$\alpha$-B3         & $ 10.0$& $-19.1$& $ 62\pm30$& $ 172\pm27$\\
H$\alpha$-C1         & $  9.9$& $-23.8$& $ 48\pm33$& $ 201\pm35$\\
H$\alpha$-C2         & $ 11.7$& $-24.8$& $ 45\pm34$& $ 178\pm39$\\
XZ-A                 & $ 14.9$& $  4.2$& $ 38\pm32$& $ 201\pm42$\\
HH 266               & $194.7$& $172.2$& $ 99\pm36$& $  43\pm21$\\
\enddata

\tablenotetext{a}
{Right ascension and declination offsets from the position of HL Tau, at 
$\alpha$(J2000)=$04^\mathrm{h}31^\mathrm{m}38\fs477$, 
$\delta$(J2000)=$+18\degr13'58\farcs86$,
for the 1999 [SII] image.}
\tablenotetext{b}
{Proper motion velocity, assuming a distance of 140 pc.}
\tablenotetext{c}
{Position angle with respect to the north direction.}
\tablenotetext{d}
{Knot HL-B has only been identified in the 1999 image. No proper motion
measurement is available.}

\end{deluxetable}

\clearpage


\begin{deluxetable}
{
l
r@{\extracolsep{1em}}
r@{\extracolsep{1.5em}}
r@{\extracolsep{1em}}
r@{\extracolsep{1.5em}}
r@{\extracolsep{1em}}
r
}
\tablewidth{0pt}
\tablecaption{
Comparison of Proper Motion Measurements in HL/XZ Tau
\label{tabhlxz}}
\tablehead{
&
\multicolumn{2}{c}{1987\tablenotemark{a}}&
\multicolumn{2}{c}{1987--1999\tablenotemark{b}}&
\multicolumn{2}{c}{1999\tablenotemark{c}}\\
\cline{2-3}
\cline{4-5}
\cline{6-7}
&
\colhead{$v_t$}&
\colhead{P.A.}&
\colhead{$v_t$}&
\colhead{P.A.}&
\colhead{$v_t$}&
\colhead{P.A.}\\
\colhead{Knot}&
\colhead{(km s$^{-1}$)}&
\colhead{(deg)}&
\colhead{(km s$^{-1}$)}&
\colhead{(deg)}&
\colhead{(km s$^{-1}$)}&
\colhead{(deg)}
}
\startdata
HL-E         &$319\pm53$ &$ 45\pm\phn4$ &$204\pm52$ &$  49\pm\phn15$ &$ 84\pm31$ &$  34\pm19$ \\
D            &$299\pm33$ &$ 38\pm\phn3$ &$178\pm52$ &$  50\pm\phn17$ &$117\pm42$ &$  44\pm14$ \\
B            &$226\pm27$ &$ 45\pm\phn3$ &$162\pm52$ &$  52\pm\phn18$ &$138\pm52$ &$  48\pm13$ \\
A            &$146\pm27$ &$ 47\pm\phn3$ &$180\pm52$ &$  46\pm\phn16$ &$114\pm43$ &$  53\pm15$ \\
HL-A         &\nodata    &\nodata       &$225\pm52$ &$  34\pm\phn13$ &$126\pm95$ &$  58\pm27$ \\
HL-B         &\nodata    &\nodata       &$ 24\pm52$ &$-134\pm124$    & \nodata   &\nodata     \\
HL-C         &\nodata    &\nodata       &$196\pm52$ &$-137\pm\phn15$ &$110\pm33$ &$ -98\pm18$ \\
HL-D1        &\nodata    &\nodata       &$121\pm52$ &$-139\pm\phn24$ &$ 49\pm40$ &$-108\pm48$ \\
HL-F         &\nodata    &\nodata       &$ 82\pm52$ &$  19\pm\phn36$ &$ 38\pm37$ &$-134\pm59$ \\
HL-G         &\nodata    &\nodata       &$133\pm52$ &$-143\pm\phn22$ &$132\pm34$ &$-119\pm16$ \\
J            &\nodata    &\nodata       &$105\pm52$ &$ -13\pm\phn28$ &$ 54\pm32$ &$ 163\pm34$ \\
H$\alpha$-B1 &\nodata    &\nodata       &$128\pm52$ &$ 144\pm\phn23$ &$ 67\pm28$ &$ 144\pm26$ \\
H$\alpha$-B2 &\nodata    &\nodata       &$ 25\pm52$ &$ -24\pm120$    &$ 44\pm31$ &$-150\pm42$ \\
H$\alpha$-B3 &$126\pm20$ &$160\pm\phn7$ &$ 49\pm52$ &$ -93\pm\phn60$ &$ 62\pm30$ &$ 172\pm27$ \\
H$\alpha$-C1 &$ 80\pm20$ &$ 97\pm25   $ &$ 44\pm52$ &$ -44\pm\phn67$ &$ 48\pm33$ &$ 201\pm35$ \\
H$\alpha$-C2 &$ 80\pm20$ &$ 97\pm25   $ &$ 62\pm52$ &$ -37\pm\phn48$ &$ 45\pm34$ &$ 178\pm39$ \\
XZ-A         &$ 80\pm33$ &$ 76\pm20   $ &$ 42\pm52$ &$  52\pm\phn71$ &$ 38\pm32$ &$ 201\pm42$ \\

\enddata

\tablenotetext{a}
{From Table 4 of Mundt et al. 1990, scaled to a distance of 140 pc.}
\tablenotetext{b}
{From positions in Mundt et al. 1990 and in this paper. An error in position of
$1''$ is assumed for all knots.}
\tablenotetext{c}
{This paper.}

\end{deluxetable}

\end{document}